\documentclass[manuscript,screen]{acmart}
\AtBeginDocument{%
  }

\setcopyright{acmlicensed}
\copyrightyear{2018}
\acmYear{2018}
\acmDOI{XXXXXXX.XXXXXXX}
\acmConference[Conference acronym 'XX]{Make sure to enter the correct
  conference title from your rights confirmation email}{June 03--05,
  2018}{Woodstock, NY}
\acmISBN{978-1-4503-XXXX-X/2018/06}

\usepackage{array}
\usepackage{listings}
\usepackage[utf8]{inputenc}
\usepackage{pdfpages}
\usepackage{spotcolor}
\usepackage{xparse}
\usepackage{blindtext}
\usepackage{morewrites}
\usepackage{wrapfig}
\usepackage{xkeyval}
\usepackage{natbib}
\usepackage{pdfpages}
\usepackage{makecell}

\usepackage{amsmath,amssymb,amsfonts}
\usepackage{algorithmic}
\usepackage{graphicx}
\usepackage{float}
\usepackage{booktabs}
\usepackage{soul, color, xcolor}
\usepackage{picinpar}
\usepackage{bbding}
\usepackage{caption}
\usepackage{pifont}
\usepackage{wasysym}
\usepackage{url}
\usepackage{hyperref}
\usepackage{float} % 在导言区
\usepackage{listings}
\usepackage{caption}
\hypersetup{
    colorlinks=true,
    linkcolor=lightblue_qing,
    filecolor=magenta,      
    urlcolor=cyan,
    citecolor= lightblue_qing,
    pdftitle={Overleaf Example},
    pdfpagemode=FullScreen,
}

\definecolor{lightblue_qing}{RGB}{47,165,212}

\newcolumntype{L}[1]{>{\raggedright\arraybackslash}p{#1}}
\newcolumntype{C}[1]{>{\centering\arraybackslash}p{#1}}

\begin{document}

%% The "title" command has an optional parameter,
%% allowing the author to define a "short title" to be used in page headers.
\title{When Blockchain Meets Crawlers: Real-time Market Analytics in Solana NFT Markets}

%%
%% The "author" command and its associated commands are used to define
%% the authors and their affiliations.
%% Of note is the shared affiliation of the first two authors, and the
%% "authornote" and "authornotemark" commands
%% used to denote shared contribution to the research.
\author{Chengxin Shen}
\affiliation{%
  \institution{School of Cyberspace Security, Hainan University}
  \city{Haikou}
  \country{China}
  \postcode{570228}
}
\authornote{Both Chengxin Shen and Zhongwen Li are co-first authors of the article.}
\author{Zhongwen Li}
\authornotemark[1]
\email{lizhongwen1230@gmail.com}
\affiliation{%
  \institution{School of Cyberspace Security, Hainan University}
  \city{Haikou}
  \country{China}
  \postcode{570228}
}

\author{Xiaoqi Li}
\affiliation{
  \institution{School of Cyberspace Security, Hainan University}
  \city{Haikou}
  % \state{Hainan}
  \country{China}
  \postcode{570228}
}
\email{csxqli@ieee.org}

\author{Zongwei Li}
\affiliation{
  \institution{School of Cyberspace Security, Hainan University}
  \city{Haikou}
  % \state{Hainan}
  \country{China}
  \postcode{570228}
}

% \author{Chengxin Shen\authornotemark[1]}
% \affiliation{
%   \institution{School of Cyberspace Security, Hainan University}
%   \city{Haikou}
%   % \state{Hainan}
%   \country{China}
%   \postcode{570228}
% }

% \author{Zhongwen Li\authornotemark[1]}
% \affiliation{
%   \institution{School of Cyberspace Security, Hainan University}
%   \city{Haikou}
%   % \state{Hainan}
%   \country{China}
%   \postcode{570228}
% }
% \email{lizhongwen1230@gmail.com}

%%
%% The abstract is a summary of the work to be presented in the
%% article.
\begin{abstract}

In this paper, we design and implement a web crawler system based on the Solana blockchain for the automated collection and analysis of market data for popular non-fungible tokens (NFTs) on the chain. Firstly, the basic information and transaction data of popular NFTs on the Solana chain are collected using the Selenium tool. Secondly, the transaction records of the Magic Eden trading market are thoroughly analyzed by combining them with the Scrapy framework to examine the price fluctuations and market trends of NFTs.
In terms of data analysis, this paper employs time series analysis to examine the dynamics of the NFT market and seeks to identify potential price patterns. In addition, the risk and return of different NFTs are evaluated using the mean-variance optimization model, taking into account their characteristics, such as illiquidity and market volatility, to provide investors with data-driven portfolio recommendations.
The experimental results show that the combination of crawler technology and financial analytics can effectively analyze NFT data on the Solana blockchain and provide timely market insights and investment strategies. This study provides a reference for further exploration in the field of digital currencies.

\end{abstract}

%%
%% The code below is generated by the tool at http://dl.acm.org/ccs.cfm.
%% Please copy and paste the code instead of the example below.
%%
% \begin{CCSXML}
% <ccs2012>
%  <concept>
%   <concept_id>00000000.0000000.0000000</concept_id>
%   <concept_desc>Do Not Use This Code, Generate the Correct Terms for Your Paper</concept_desc>
%   <concept_significance>500</concept_significance>
%  </concept>
%  <concept>
%   <concept_id>00000000.00000000.00000000</concept_id>
%   <concept_desc>Do Not Use This Code, Generate the Correct Terms for Your Paper</concept_desc>
%   <concept_significance>300</concept_significance>
%  </concept>
%  <concept>
%   <concept_id>00000000.00000000.00000000</concept_id>
%   <concept_desc>Do Not Use This Code, Generate the Correct Terms for Your Paper</concept_desc>
%   <concept_significance>100</concept_significance>
%  </concept>
%  <concept>
%   <concept_id>00000000.00000000.00000000</concept_id>
%   <concept_desc>Do Not Use This Code, Generate the Correct Terms for Your Paper</concept_desc>
%   <concept_significance>100</concept_significance>
%  </concept>
% </ccs2012>
% \end{CCSXML}

% \ccsdesc[500]{Do Not Use This Code~Generate the Correct Terms for Your Paper}
% \ccsdesc[300]{Do Not Use This Code~Generate the Correct Terms for Your Paper}
% \ccsdesc{Do Not Use This Code~Generate the Correct Terms for Your Paper}
% \ccsdesc[100]{Do Not Use This Code~Generate the Correct Terms for Your Paper}

%%
%% Keywords. The author(s) should pick words that accurately describe
%% the work being presented. Separate the keywords with commas.

%%
%% This command processes the author and affiliation, and title
%% information and builds the first part of the formatted document.
\maketitle

\section{Introduction}\label{1}
\

With the rapid development of blockchain technology, nonhomogenized tokens (NFTs), as one of its important applications, have made significant progress in the fields of art, gaming, entertainment, etc \cite{mougayar2016business}. NFTs have quickly gained the attention of a large number of investors and collectors in the market because of their uniqueness and irreplaceability as digital artworks and virtual assets. However, the high volatility of the NFT market and the illiquidity of the market make investors face greater risks\cite{zhang2022authros}. Therefore, understanding the market trend and formulating reasonable investment strategies using effective data analysis methods has become an important research topic in the current field of NFT investment \cite{cr2025blockchain}.

\begin{figure}[H]
    \centering
    \includegraphics[width=0.6\linewidth]{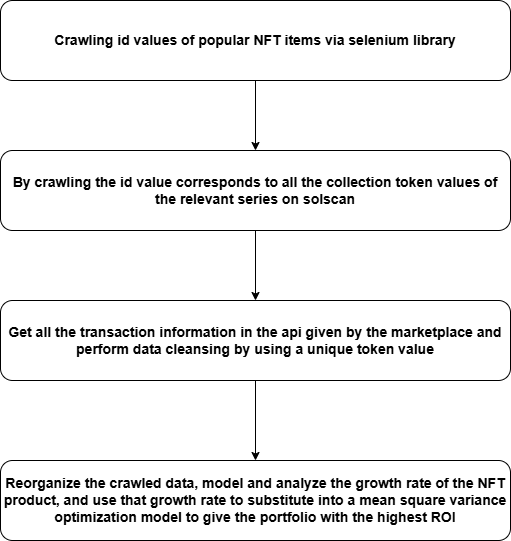}
    \caption{ The Main Idea Steps in This Paper}
    \label{fig:1}
\end{figure}

This paper aims to design and implement a web crawler system based on the Solana blockchain to automatically collect the transaction data of popular NFT projects in the Solana chain \cite{wu2025atomicity, xiao2025parallelizing, aguilarSmartContractFamilies2024} and combine time series analysis techniques with a mean-variance optimization model to help investors make more data-driven investment decisions \cite{pierro2022can, arceriSoundConstructionEVM2024, ayubSoundAnalysisMigration2024}. The study uses Selenium and Scrapy frameworks to crawl the historical transaction data on the Magic Eden platform \cite{caiEnablingCompleteAtomicity2024}, applies time series analysis to explore the dynamic changes of the NFT market \cite{chenDemystifyingInvariantEffectiveness2024}, uses a mean-variance optimization model to evaluate the risk and return of different NFTs \cite{wang2024ContractsentryStaticAnalysis, sun2025FIRESmartContract}, and ultimately recommends reasonable NFT portfolios for investors \cite{auksutis2024investicinio, grossmanPracticalVerificationSmart2024}. The main idea steps in this paper are shown in Fig. \ref{fig:1}.
\section{Background}\label{2}
\

Since the introduction of Bitcoin, blockchain technology has developed rapidly, and its application fields have expanded from cryptocurrency to supply chain, healthcare, finance, etc\cite{zhu2024sybil}. Solana, as a high-performance blockchain platform, has made great strides in the field of NFTs and decentralized applications (DApps) because of the advantages of fast transaction speed, low cost, etc \cite{schletz2020blockchain}. NFT, as a unique digital asset, has been widely used in industries such as art, gaming, etc., but its market volatility and uncertainty have also prompted the urgent need to deeply analyze its market dynamics. Other industries, but the volatility and uncertainty of its market have also prompted an urgent need for in-depth analysis of its market dynamics \cite{kubinska2023behavioral}.

With the rapid development of the NFT market, real-time and accurate data collection has become increasingly important \cite{azaria2016medrec, hanOSwapPreservingAtomicity2026, heCodeNotNatural2024}. The inefficiency of traditional manual data collection methods has led to the widespread use of automated tools such as Selenium and Scrapy \cite{huangAdvancingWeb302024, jiaoSurveyEthereumSmart2024}. These tools speed up data collection, improve analytical efficiency, and provide investors with more accurate market forecasts and strategies\cite{zhong2023sybil, kumarVulnerabilitiesSmartContracts2024}. In the highly volatile NFT market, methods such as time series analysis and mean variance optimization are widely used to assess the risk and return of different NFT assets to help optimize portfolios \cite {chang2020blockchain, liASTRODetectingAccess2025, liDemoEnhancingSmart2024}.

\subsection{Selenium}\label{2.1}
\

Selenium is an open-source automated testing tool primarily used for automated testing of web applications. It supports multiple browsers, including Chrome, Firefox, Safari, and Internet Explorer, and is capable of executing automated tests on a variety of platforms and environments. The core functionality of Selenium is to automate the operation of web browsers through a toolset that utilizes advanced techniques to remotely control browser instances and simulate user-browser interactions \cite{garcia2021automated}.
The main advantage of Selenium is its ability to simulate common actions performed by end-users, such as text input, drop-down selections, checking checkboxes, and clicking on page links. In addition, Selenium provides a variety of advanced controls, such as mouseover and JavaScript execution. These features make Selenium not only a powerful tool for web automation testing but also an important tool in data crawling, especially for crawling dynamic web content. In contrast, Scrapy mainly deals with static HTML pages and has limited support for JavaScript-rendered content \cite{chang2023blockchain}.

In this paper, for the characteristics of websites such as solscan.io, whose web content is mostly loaded on the client side and lacks open public API interfaces, Selenium is used to simulate browser operations for data crawling\cite{liu2025sok}. This approach can effectively deal with dynamically generated page content by JavaScript and is therefore widely used in this paper.

\subsection{Scrapy}
\

Scrapy is an open-source framework for web crawling and data extraction, developed primarily in Python. It provides developers with a complete set of tools to support extracting data from websites, processing it, and storing it in a specified format\cite{zhang2017android}. Scrapy uses a Twisted-based non-blocking (asynchronous) web framework to handle network communication, resulting in efficient download performance. Its key features include simplicity, flexibility, scalability, and high performance \cite{ze4588309characterizing}.  The core components of Scrapy are shown in the Tab~\ref{tab:scrapy_components}. The advantages of Scrapy in blockchain data crawling are mainly reflected in the following aspects.

\begin{table}[h]
\centering
\caption{Scrapy Framework Components}
\label{tab:scrapy_components}
\begin{tabular}{@{} c c @{}}
\toprule
\textbf{Component} & \textbf{Description} \\ 
\midrule
Spiders & Crawler defines the behavior of crawling the site and page parsing methods. \\
Items & Items, defining the data structure of the crawl. \\
Item Pipelines & Pipelines, which process the crawled data. \\
Downloader & Downloader, responsible for fetching web content. \\
Engine & The engine, which controls the flow of data between components. \\
Scheduler & The scheduler, which manages the requests of the crawl. \\
\bottomrule
\end{tabular}
\end{table}

\begin{enumerate}
    \item  Data crawling efficiency: Scrapy is based on asynchronous network communication, which can efficiently handle a large number of data requests. Given the fast block-out speed of the Solana chain and the huge amount of data, the Scrapy framework can efficiently crawl the data on the Solana chain.
    \item Data processing and analysis: through Scrapy's Items and Item Pipelines components, developers can effectively structure, clean, and store crawled data for subsequent data analysis.
    \item  Scalability:  Scrapy allows developers to customize the crawler and pipelines, and to tailor the crawler logic and data processing flow to the specific needs of the Solana blockchain\cite{chen2018system}.
\end{enumerate}

Therefore, in crawling and analyzing for NFT market prices, Scrapy not only improves the efficiency of data crawling and processing but also helps developers to customize and optimize according to specific needs through its flexible architecture and strong community support \cite{di2017blockchain}.
\vspace{-2ex}
\subsection{Time Series Analysis}
\

Time series analysis is a statistical technique that is specifically used to analyze data points in chronological order. In this type of analysis, data exists in equally spaced time series, e.g., minute-by-minute, day-by-day, month-by-month, and so on. Time series data is characterized by autocorrelation, i.e., data at one point in time may be correlated with data at points in time before and after it\cite{mao2024scla}. The goal of time series analysis is to understand and model the inherent structure and patterns of data over time for effective forecasting and decision-making \cite{kirchgassner2012introduction}.
Considering that NFT products have low liquidity and do not have significant price changes in the market, specific time series analysis models (e.g., ARIMA or seasonal decomposition, etc.) have not been directly applied in this paper, but rather, we have focused on how to handle and utilize time series data, especially in evaluating the performance of NFTs and performing portfolio optimization\cite{zou2025malicious, priftiSmartContractVulnerability2024, suDiSCoDecompilingEVM2025, wangContractCheckCheckingEthereum2024}. 

The actual algorithm used is not the standard approach in traditional time series analysis, but it is based entirely on the important idea of time series data. In this paper, we try to give different weights to the data at different time points by considering weighted returns over time intervals. For prices \( P_t \)  and \( P_{t+1} \) at any two consecutive time points \( t \) and \( t+1 \) the simple rate of return \( R \) can be expressed as eq.~\eqref{eq: standardized method}., which is the standard method for calculating the change in price over each time interval \cite{kim2022ultra}.
\begin{equation}
R = \frac{P_{t+1} - P_t}{P_t}
\label{eq: standardized method}
\end{equation}

Considering the specific number of seconds in each time interval  $\Delta t$, we adjust Eq.~\eqref {eq: standardized method}, and the adjusted interval return is shown in Eq.~\eqref {eq: method of adjustment}. $\Delta t$ is the time difference in seconds between two price points. The purpose of this formula is to adjust the simple return to an equivalent compound return per second, thus allowing for fair comparisons at different time intervals\cite{bu2025smartbugbert}.
 
\begin{equation}
R_{\text{adjusted}} = (1 + R)^{\frac{1}{\Delta t}} - 1
\label{eq: method of adjustment}
\end{equation}

By multiplying the adjusted returns for all intervals, the overall weighted return for the entire observation period can be calculated, as shown in Eq.~\eqref {eq: total method}. Where \( n \) is the number of time intervals and 
$R_{\text{adjusted},i}$
is the adjusted return for the \( i \)-th interval. This formula takes into account the compounding effect of returns within each interval and provides a comprehensive assessment of asset performance over the entire period\cite{li2021hybrid}.

\begin{equation}
R_{\text{total}} = \prod_{i=1}^{n} (1 + R_{\text{adjusted},i}) - 1
\label{eq: total method}
\end{equation}

\subsection{Mean-variance Optimization}
\

Mean-Variance Optimization (MVO) is a fundamental component of modern portfolio theory, proposed by Harry Markowitz in 1952, which won him the Nobel Prize in Economics. The core idea of this theory is that portfolio selection should take into account not only the expected return (mean) but also the risk (variance or standard deviation), and seek to maximize the expected return at a given level of risk or minimize the risk at a given level of expected return \cite{kalayci2019comprehensive, wangEfficientlyDetectingReentrancy2024, wangEmpiricalAnalysisSmart2026, weiSurveyQualityAssurance2024}.
The goal of MVO is to find the optimal asset allocation, i.e., the weights of the different assets in the portfolio, to form an “efficient frontier”, where each point on the frontier represents a portfolio that is optimized in terms of risk and return\cite{li2017discovering}. In this paper, the objective of MVO is to determine the optimal asset weights to maximize the Sharpe ratio, i.e., to solve the optimization problem mentioned in Eq.~\eqref {eq: maximize}:

\begin{equation}
\text{Maximize} = \frac{E[R_p] - R_f}{\sigma_p}
\label{eq: maximize}
\end{equation}

Where $E[R_p]$ is the expected return of the portfolio,   $R_f$ is the risk-free rate, and $\sigma_p$ is the standard deviation (risk) of the portfolio. In practice, it is possible to transform this optimization problem into minimizing the negative Sharpe ratio, since the minimum value rather than the maximum value is sought when using “scipy.optimize.minimize” \cite{kourtis2015stability}.
\vspace{-2ex}
\subsection{Anti-crawler}
\

As the amount of data on the Internet grows, data (e.g., text, images, videos, etc.) provided by websites and online services becomes an important resource for organizations. To protect data from malicious crawling, many websites employ anti-crawler techniques such as FingerprintJS, Captchas, and dynamic interactive validation, which effectively limit access to automated tools \cite{10090174}.
However, anti-crawler techniques also make crawler development more difficult. Crawler developers need to address these tactics, especially since tools like Selenium can bypass these safeguards and reduce the risk of detection by simulating real user behavior \cite{zhangEVMShieldInContractState2024, zhangInferringLikelyCountingrelated2025}. Selenium can efficiently crawl dynamically loaded JavaScript pages, further enhancing crawling efficiency \cite{wan2019pathmarker, zhuSurveySecurityAnalysis2024, boi2024VulnHuntGPTSmartContract, hu2023LargeLanguageModelPowered, wei2025AdvancedSmartContract}.
\subsubsection{Remove Navigator.Webdriver Flag}
\

\texttt{Navigator.Webdriver} is a flag that indicates whether the browser is being controlled by an automation tool such as Selenium\cite{li2025scalm}. When the \texttt{Webdriver} is launched with Selenium, the browser displays “Chrome is being controlled by automated test software”, which means that the website can use this flag to know that the visitor is an automated program and not a human user. This flag is set to true, allowing websites to recognize automated visits with simple detection methods\cite{wang2024smart}. 
In order to prevent automated tools, a site may take steps such as passing Cloudflare's five-second human verification.
This flag typically appears when parameters such as \texttt{\texttt{——enable-automation}}, \texttt{\texttt{——headless}}, \texttt{\texttt{——remote-debugging-port}}
are enabled. 

The analytical tools available on the application website are usually shown in Listing~\ref{lst:webdriver-check}.
Since Boolean checking is simpler, it is possible to use the Listing~\ref{lst: webdriver settings} for Chrome WebDriver to change its characteristics at runtime so that Chrome does not recognize that Selenium is being used \cite{garcia2021automated}.
% Copyright 2017 Sergei Tikhomirov, MIT License
% https://github.com/s-tikhomirov/solidity-latex-highlighting/

%\usepackage{listings, xcolor}

\definecolor{verylightgray}{rgb}{.97,.97,.97}

\lstdefinelanguage{Solidity}{
	keywords=[1]{anonymous, assembly, assert, balance, break, call, callcode, case, catch, class, constant, continue, constructor, contract, debugger, default, delegatecall, delete, do, else, emit, event, experimental, export, external, false, finally, for, function, gas, if, implements, import, in, indexed, instanceof, interface, internal, is, length, library, log0, log1, log2, log3, log4, memory, modifier, new, payable, pragma, private, protected, public, pure, push, require, return, returns, revert, selfdestruct, send, solidity, storage, struct, suicide, super, switch, then, this, throw, transfer, true, try, typeof, using, value, view, while, with, addmod, ecrecover, keccak256, mulmod, ripemd160, sha256, sha3}, % generic keywords including crypto operations
	keywordstyle=[1]\color{blue}\bfseries,
	keywords=[2]{address, bool, byte, bytes, bytes1, bytes2, bytes3, bytes4, bytes5, bytes6, bytes7, bytes8, bytes9, bytes10, bytes11, bytes12, bytes13, bytes14, bytes15, bytes16, bytes17, bytes18, bytes19, bytes20, bytes21, bytes22, bytes23, bytes24, bytes25, bytes26, bytes27, bytes28, bytes29, bytes30, bytes31, bytes32, enum, int, int8, int16, int24, int32, int40, int48, int56, int64, int72, int80, int88, int96, int104, int112, int120, int128, int136, int144, int152, int160, int168, int176, int184, int192, int200, int208, int216, int224, int232, int240, int248, int256, mapping, string, uint, uint8, uint16, uint24, uint32, uint40, uint48, uint56, uint64, uint72, uint80, uint88, uint96, uint104, uint112, uint120, uint128, uint136, uint144, uint152, uint160, uint168, uint176, uint184, uint192, uint200, uint208, uint216, uint224, uint232, uint240, uint248, uint256, var, void, ether, finney, szabo, wei, days, hours, minutes, seconds, weeks, years},	% types; money and time units
	keywordstyle=[2]\color{teal}\bfseries,
	keywords=[3]{block, blockhash, coinbase, difficulty, gaslimit, number, timestamp, msg, data, gas, sender, sig, value, now, tx, gasprice, origin},	% environment variables
	keywordstyle=[3]\color{violet}\bfseries,
	identifierstyle=\color{black},
	sensitive=false,
	comment=[l]{//},
	morecomment=[s]{/*}{*/},
	commentstyle=\color{gray}\ttfamily,
	stringstyle=\color{red}\ttfamily,
	morestring=[b]',
	morestring=[b]"
}

\lstset{
	language=Solidity,
	%backgroundcolor=\color{verylightgray},
	extendedchars=true,
	basicstyle=\normalsize\ttfamily,
	showstringspaces=false,
	showspaces=false,
	numbers=left,
	numberstyle=\normalsize,
	numbersep=9pt,
	tabsize=2,
	breaklines=true,
	showtabs=false,
	captionpos=b
}

\begin{figure}[H]
\begin{lstlisting}[caption={Check on the Server for Automated Tool Access}, label={lst:webdriver-check}]
{
  var isAutomated = navigator.webdriver;
  if(isAutomated){
    blockAccess();
  }
}
\end{lstlisting}
\end{figure}

\begin{lstlisting}[caption={Changing Webdriver Settings},label={lst: webdriver settings}]
{
  chrome_options.add_experimental_opton('useAutomationExtension',False)
  chrome_options.add_experimental_opton("excludeSwitches",['enable-automation'])
}
\end{lstlisting}
\vspace{-2ex}

\subsubsection{Obfuscate JavaScript in Browser Drivers}
\

When opening \texttt{chromedriver.exe}  in a text editor and locating it at about line 4000, a characteristic piece of JavaScript code can be found, which is run when using Selenium. This code is often used by anti-crawler tools such as \texttt{FingerprintJS},\texttt{Imperva} (formerly Distil Networks), and Google Captcha to detect automated access behavior\cite{zhong2023sybil}. Since this JavaScript is directly exposed, it can be bypassed by modifying its variable names, but it is important to keep the variable names consistent in length, as this may cause Selenium to crash \cite{bielova2013survey}.
The most critical of these variables is named \texttt{\$ cdc\_asdjflasutopfhvcZLmcfl\_ \_}, which is the signature string that human authentication systems typically search for. Some of the detection mechanisms can be circumvented by replacing them with another variable name of the same length, randomly generated (e.g. \$btlhsaxJbTXmBATUDvTRhvcZLm\_)\cite{li2024detecting}. The anti-anti-crawler procedure is as follows:
\begin{enumerate}
    \item Open chromedriver.exe with a text editor.
    \item Use the shortcut Ctrl+F to search for the key variable name \texttt{\$cdc\_asdjflasutopfhvcZLmcfl\_}.
    \item Replace this variable name with any string of the same length.
    \item After saving the file, you can use it as a new driver to improve the success rate of page data crawling.
\end{enumerate}
\subsubsection{Use a Proxy to Change IP Address to Circumvent a Ban}
\

\begin{figure}[ht]
    \centering
    \includegraphics[width=0.55\linewidth]{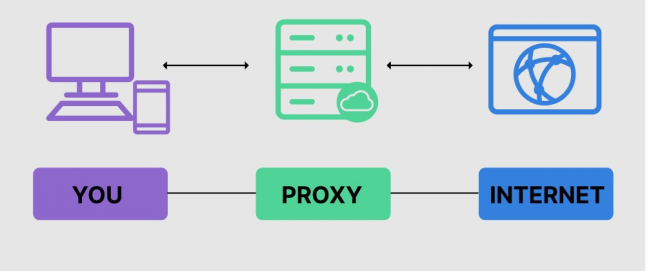}
    \caption{Proxy Server Role Schematic}
    \label{fig: Proxy Server Role Schematic}
\end{figure}

Some companies that provide web security services, such as Cloudflare, analyze visitor behavioral characteristics, including request frequency, page access order, and HTTP request header information, through their edge nodes to identify automated access behavior. If the behavior of an IP address deviates significantly from normal user patterns, the system may limit or even block access requests from that IP\cite{bu2025enhancing}. Therefore, in order to crawl a large number of pages, it is common practice to distribute requests across multiple hosts or to use a proxy server to hide the true source of access. The principle of proxy is that the crawler request is first sent to the proxy server, and then the proxy forwards the request to the target website, and the source IP seen by the website is the address of the proxy server. By building IP pools and rotating proxies, crawlers can simulate users from different geographic locations and devices, thus avoiding the risk of blocking due to frequent visits from a single IP \cite{kreimer2006censorship}.

In addition, IP rotation helps break up the behavioral fingerprint of requests, making automated access harder to detect. Even if the original IP is blocked, you can continue to access the target site by changing proxies, thus bypassing IP or geographic location-based access restrictions. Listing~\ref{lst: Proxy server} demonstrates how to use a proxy server to access, at the same time, Fig.~\ref {fig: Proxy Server Role Schematic} illustrates the process of proxy server action\cite{zhong2023tackling}.

\begin{lstlisting}[caption={Proxy Server Access},label={lst: Proxy server}]
{
  #Add Proxy
  option = webdriver.ChromeOptions()
  option.add_argument('prcxy-server=106.122.8.54:3128')
  #open Browser
  browser = webdriver.chrome(executable_path='chromedriver.exe',options=option)
}
\end{lstlisting}

Although there are a large number of free proxies on the Internet (e.g. Proxy 66, Proxy360, Goubanjia, etc. provide free http proxies), the proxies they provide are also used by a large number of other tagged web crawlers at the same time, which leads to the fact that nowadays, many free IPs are also tagged as bots. Pools are built with far fewer HTTPS proxies \cite{degefa2022mes}.
\

\subsubsection{Use Cookies to Maintain Sessions and Bypass Authentication }
\

Cookies are small pieces of data that a server sends to a user's browser and stores locally, which are automatically carried by the browser on subsequent visits to the same server to maintain the session state. Typically, cookies are used to recognize multiple requests from the same user, allowing them to jump around a site without having to log in again \cite{mundada2016half}.
In web crawlers, cookies are used to simulate normal user behavior. When a crawler visits a target site for the first time, the server may set cookies to identify the user. At this point, the crawler needs to capture and save these cookies, and then carry and send them in subsequent requests to simulate an ongoing session. Most crawler frameworks, such as Python's requests library, have the ability to handle cookies automatically.

In this way, crawlers are able to access pages that require a login or a specific user state to access. In addition, some sites use cross-site cookies to detect a user's page flow path and prevent the user from accessing restricted pages directly via URL. Crawlers that correctly emulate cookie behavior can also bypass these access restrictions \cite{dacosta2012one}.

\section{Methods}\label{3}
\subsection{Crawling top 50 stock NFT series}
\

Use the selenium library to crawl the id value of the popular NFT series on scans, Listing~\ref{lst: Crawl} shows how to crawl the id value of the popular NFT, this operation is convenient to follow up on the series of all the collections of the token to crawl while using regular expressions to match, the specific rules are shown below, Listing~\ref{lst: Regular Expressions Match}  shows the regular expression to match the required fields.
\vspace{5ex}

\begin{lstlisting}[caption={Crawling the Id Values of Popular NFTs},label={lst: Crawl}]
{
  With open("Seriesid_example.json",'w') as f:
      NFTSeriesId = crawLNFSeriesInSolscan()
      json.dump(NFTSerisesId,fp=f)
}
\end{lstlisting}

\vspace{-2ex}

\begin{figure}[H]
\begin{lstlisting}[caption={Regular Expressions Match the Required Fields },label={lst: Regular Expressions Match}]
{
  url = "https://pro-api.solscan.io/v1.0/public/nft/collection/overview?sort_by=volume&offset=0&limit=50&range=30&sort_order=desc&cluster=" 
  NFTSeriesList = [] 
  try:
      browser.get(url)
      time.sleep(2)
      tmp_collectionid = re.findall(r'"collection_id":"[A-Za-z0-9]*"', browser.page_source) tmp_volume = re.findall(r'\"volume\":[\.0-9]*\,', browser.page_source)
      tmp_name = re.findall(r'"collection_name":\s*"(.*?)"', browser.page_source)
}
\end{lstlisting}
\end{figure}
\vspace{-2ex}

\begin{enumerate}
    \item \verb|‘\"collection_id\":\"’| : matches the literal string \verb|‘"collection_id":"’| Use \verb|‘\’| to escape quotes \verb|‘"’|.

    \item \verb|‘[A-Za-z0-9]*’|: matches any number of letters (upper or lower case) and numbers. \verb|‘*’| means match the preceding character (in this case any letter or number) 0 or more times.

    \item The \verb|‘tmp_collectionid’| expression matches strings of the form \verb|‘"collection_id":"ABCD123"’|, excluding closing quotes.

    \item \verb|‘\"volume\":’|: matches the literal string \verb|‘"volume":’|. Similarly, \verb|‘\’| is used to escape the single quote \verb|‘"’|.

    \item \verb|‘[\.0-9]*’|: matches any number of decimal points or digits. The decimal point needs to be escaped (\verb|\.|), because in regular expressions \verb|.| has a special meaning (matches any single character).

    \item \verb|‘\,’|: matches the literal comma \verb|‘,’|. The comma is also escaped, although escaping the comma is not necessary in this case.

    \item \verb|‘tmp_volume’|: The entire expression matches strings of the form \verb|‘"volume":12345,’|, where the numeric portion can be any positive integer or floating-point number.

    \item \verb|‘\"collection_name\":\s*\"’|: matches the literal string \verb|‘"collection_name":’| followed by any number of whitespace characters (\verb|‘\s*’|).
    \item \verb|'(.*?)'|: capture group to match and capture any number of characters 
(\verb|'.*'|) until the next specified pattern is encountered 
- the next single quote \verb|'| after \verb|.*| 
makes the match non-greedy, meaning it will match as few characters as possible to fulfill the pattern.

    \item The \verb|‘tmp_name’| expression matches strings shaped like \verb|‘"collection_name": "Some Collection Name"’| and extracts \verb|‘Some Collection Name’| by capturing the group.
\end{enumerate}
\

After de-weighting the operations in Listing~\ref{lst: Crawling for Popular NFTs}, a JSON file corresponding to the series and ID is obtained, which contains the corresponding ID value of the NFT collection stored on the Solscan server of the website and thus facilitates the next crawling operation.

\begin{lstlisting}[caption={ The Process of Id Crawling for Popular NFTs},label={lst: Crawling for Popular NFTs}]
{
   "collection_id": "4Q2C5S930M9c9e96b...",
   "collection_name": "Frogman",
   "floor_price": 807080.2,
   "last_trade_time": "1711308573",
   "total_attributes": 6969,
   "marketplace": "Magic Eden",
   "volume": 3000000,
   "id": "A2MxSTGcBGTRyK97K..."
}
\end{lstlisting}
\subsection{Crawling token values}
\

Since crawling the site directly via Scrapy returns an HTTP 403 status code, which is not the case with browser emulation using the Selenium library, we chose to use Selenium to obtain the token value. Note that you must use a proxy to access the site, otherwise, Cloudflare's anti-bot authentication mechanism may be triggered, causing access to fail.

The crawling process consists of the following steps: first, the page is scrolled to the bottom by calling JavaScript; second, the browser is set to display 50 items of data per page; and image loading is disabled to reduce network traffic and speed up the page loading. Subsequently, XPath is used to accurately match the target content in the webpage, and combined with the page-turning mechanism, the required data is continuously crawled. The specific code formulation is as follows: Listing~\ref{lst: Corresponding NFT Product Token}.
\begin{lstlisting}[caption={Continuously Crawl the Corresponding NFT Product Token by Controlling the Page Flip},label={lst: Corresponding NFT Product Token}]

   #Locate the element in the hover box and click to view 50 pages
   . element_to_click = browser.find_element(By.CLASS_NAME," ant-select") element_to_click.click()
   time.sleep(1)
   browser.execute_script("window.scrollTo(8, document.body.scrollHeight);") 
   select_item = browser,find_element(By,XPATH,
   "//div[contains(@class,'ant-select-item-option-content') and text()='50']*)
   select_item.click() 
   time.sleep(2)
\end{lstlisting}

\begin{enumerate}
    \item \lstinline|//div|: Start by searching for all \verb|<div>| elements from any position in the document. \verb|//| is a wildcard in XPath, indicating the selection of child nodes at any depth from the current node.
    
    \item \lstinline|[contains(@class, 'ant-select-item-option-content')]|: This predicate is used to further narrow down the selected \verb|<div>| elements to those that meet specific conditions. 
    
    Here, the expression \verb|contains(@class, 'ant-select-item-option-content')| looks for \verb|<div>| elements whose \verb|class| attribute contains the string \verb|ant-select-item-option-content|. \verb|@class| refers to the \verb|class| attribute of the element, and the \verb|contains()| function checks whether the attribute value includes the given substring.

    \item \lstinline|[text()='50']|: This predicate further restricts the \verb|<div>| elements to those with text content exactly equal to \verb|50|. The \verb|text()| function retrieves the text content of the element and compares it to \verb|50|.
\end{enumerate}

\begin{figure}[H]
    \centering
    \includegraphics[width=1\linewidth]{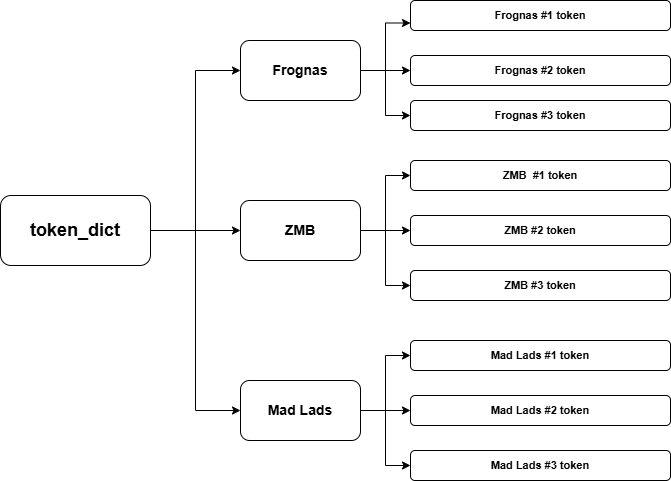}
    \caption{Crawled NFT Collection Storage Structure}
    \label{fig:13}
\end{figure}

Combined, the entire \texttt{XPath} expression searches for \verb|'<div>'| elements that satisfy all the following conditions: located anywhere in the document; have a \verb|' class'| attribute containing the string \verb|'ant-select-item-option-content'|; and have text content exactly equal to \verb|'50'|.The structure of the JSON storage of the crawled data is shown in  Fig. \ref{fig:13}. It is also possible to see the general picture of the NFT by reading it using the Python console.
\newpage
\subsubsection{Exception Handling in Token Crawling}
\

In order to prevent the "Next Page" button of the target webpage from loading in time during the crawling process, a waiting mechanism is set in the crawling logic, otherwise, it will be judged that the crawling of the current series has ended. The overall crawling process is to try to click the "next page" button on the web page in a loop, and the following processing is carried out to deal with the abnormal situations that may occur:
\begin{enumerate}
\item  \verb|TimeoutException|: The \verb|'TimeoutException'| exception is thrown when \verb|'WebDriverWait'| fails to find a clickable element within the specified time (20 seconds). This means that the button is not displayed on the page, most likely because the page is not fully loaded yet. In this case, the loop is broken (\verb|'break'|), and no further attempts are made to click the button.

\item  \verb|StaleElementReferenceException|: attempting to interact with an element when it no longer appears on the DOM, or if the element is now attached to a different part of the DOM, will throw a \verb|'StaleElementReferenceException'| exception. This usually happens after the page has been partially or fully refreshed. The loop is also interrupted in this case.

\item  \verb|ElementClickInterceptedException|: The \verb|'ElementClickInterceptedException'| exception is thrown when the element that was attempted to be clicked on is obscured by another element. This can be due to a pop-up or other UI element on the web page covering the button in the attempted click. The exception is caught by calling a JavaScript function to attempt to click the button again after scrolling to the bottom of the page. If the \verb|'ElementClickInterceptedException'| exception is thrown again, the loop is interrupted, and it is assumed that all current items have been crawled.

\item  \verb|'finally'| clause: the \verb|'finally'| clause is always executed, regardless of whether the previous \verb|'try'| and \verb|'except'| blocks are successfully executed. In this clause, the code first extracts some information from the page source code via regular expressions and adds this information to a cumulative list. It then tries to prepare for the next button click by scrolling to the bottom of the page.

\item  \verb|Clicking the button|: if no exception is encountered during the attempt, \verb|'button.click()'| is executed to try to click the button. If \verb|'ElementClickInterceptedException'| occurs while clicking, the loop is interrupted, i.e., if the last side is reached and there is no next page for the `button` symbol, at this point the exception is thrown and the loop is exited for the next collection series to be crawled.
\end{enumerate}

The exception handling mechanism is designed to improve the robustness of the crawler so that, in the face of network delay, dynamic loading, page pop-ups, and other practical problems, it can stably crawl the required data and automatically complete the state judgment and error recovery. The specific crawling stop and exception handling code is in Listing~\ref{lst: Crawl Stop and Exception Handling}.

\subsection{Crawling NFT Product Transaction Information  }
\

In the website under the Solscan domain, this pa successfully captured the token value of the target NFT. The token is a unique address identifier that corresponds to the initial creation of each NFT. The collected token values for each NFT family were then used to access the API provided by the main NFT marketplace Magic Eden via the Scrapy framework. 
By calling the API, detailed information about each NFT project, such as transaction history, can be extracted.

\newpage

\begin{lstlisting}[caption={Continuously Crawl the Corresponding NFT Product Token by Controlling the Page Flip},label={lst: Crawl Stop and Exception Handling}]
   while True:
      try:
        button = WebDriverWait(browser,20).until(
              EC.element_to_be_clickable((By.XPATH,'//*[@id="rc-tabs-0-panel-default"]/div[3]/button[2]/span'))
          )
      except TimeoutException as e:
          break
      except StaleElementReferenceException as e:
          break
      except ElementClickInterceptedException as e:
    browser.execute_script("window.scrollTo(0,document.body.scrollHeight);") 
            try:
                button = WebDriverWait(browser,20).until(
                EC.element_to_be_clickable((By.XPATH, '//*[@id="rc-tabs-0-panel-default"]/div[3]/button[2]/span'))
                )
                except ElementClickInterceptedException:
                    break
    finally:
        limit50tokenlist = re.findall(r'<a href="/token/([a-zA-Z0-9]*)"', browser.page_source) tmplimit50token = list(set(limit50tokenlist)) NFTSeriesTokens += tmplimit50token
 
        browser.execute_script("window.scrollTo(0,document.body.scrollHeight);")
\end{lstlisting}
\vspace{-1em}

\begin{figure}[H]
    \centering
    \includegraphics[width=0.9\linewidth]{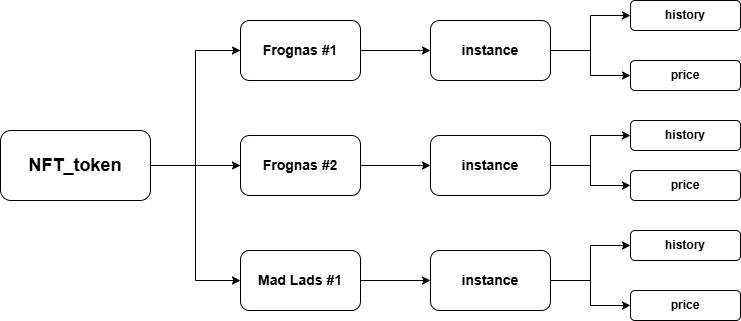}
    \caption{Storage Model Definition for NFT Crawl Results}
    \label{fig:15}
\end{figure}

To manage the data in a structured manner, this paper defined the following data models in the Scrapy project. The storage model for NFT crawling results is defined as shown in Fig. \ref{fig:15}. Each \lstinline|NftInstanceItem| contains the fields shown in the Tab ~\ref{tab:nft_instance_structure}. The Listing~\ref{lst:item_definitions} shows the definition of the item's data structure. \text{\lstinline|NftCollectionItem|}: Defined in the
\lstinline | items.py | file to organize the overall information of each NFT collection. \text{\lstinline|NftInstanceItem|}: Represents an individual NFT item and includes its transaction records.

\begin{table}[H]
\vspace{-4ex}
\centering
\caption{Structure of \texttt{NftInstanceItem}}
\vspace{-2ex}
\begin{tabular}{ll}
\toprule
\textbf{Field} & \textbf{Description} \\
\midrule
\texttt{instance} & The unique identifier of the NFT. \\
\texttt{token} & The Solana address bound to this instance. \\
\texttt{series\_name} & The name of the corresponding collection or series. \\
\texttt{history} & A list of historical transaction events associated with this NFT. \\
\texttt{price} & A list of prices corresponding to the events in \texttt{history}. \\
\bottomrule
\end{tabular}
\label{tab:nft_instance_structure}
\end{table}

\vspace{-3ex}

\begin{lstlisting}[language=Python, caption={Definition of NftInstanceItem and NftCollectionItem Classes}, label={lst:item_definitions}]
import scrapy
class NftInstanceItem(scrapy.Item):
    # Historical prices and their corresponding timestamps of the NFT instance
    history_time = scrapy.Field()
    price = scrapy.Field()
class NftCollectionItem(scrapy.Item):
    # Dictionary of NftInstanceItem
    nft_token = scrapy.Field()
    nft_name = scrapy.Field()
    nft_instances = scrapy.Field()
\end{lstlisting}

\subsubsection{Spider Structural Design}
\

Building well-structured \texttt{Spider} modules in the Scrapy framework not only improves the overall development efficiency of the project but also offers significant advantages in terms of maintainability, scalability, performance optimization, and readability:
\begin{itemize}
    \item Maintainability: A well-structured \texttt{Spider} module improves code comprehensibility. When an error or exception occurs during a crawling task, a clean structure allows for quick identification and resolution of issues, thereby reducing maintenance costs.
    \item Scalability: During project development, it is often necessary to extend the data sources or adapt to changes in the target website's structure. A modular and well-architected Spider facilitates the addition of new features without disrupting existing functionality and allows logic to be reused across different Spiders.
    \item Performance optimization: Proper structural design helps in controlling the frequency of network requests and avoids redundant data processing, thereby improving crawler execution efficiency. Centralized user-agent management and request interval settings also help to mimic legitimate user behavior and reduce the risk of being blocked by the target website.
    \item Readability: A concise and logical codebase improves overall readability. New contributors can more easily understand and work with the code, thus enhancing team collaboration and development efficiency.
\end{itemize}

\subsubsection{Core Method Implementation}
\

\begin{figure}[H]
    \centering
    \includegraphics[width=0.6\linewidth]{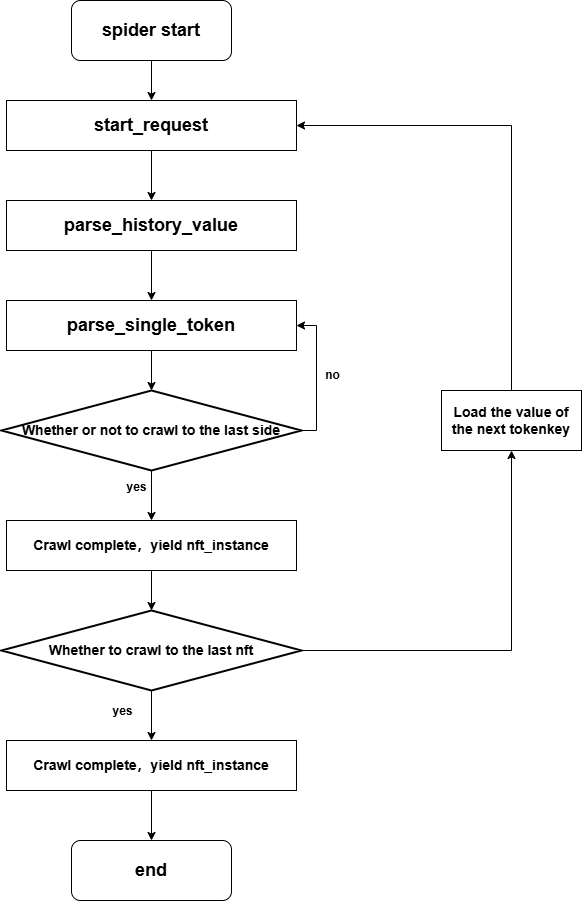}
    \caption{Scrapy crawl transaction information process}
    \label{fig:21}
\end{figure}

The flow of the crawling process using the Scrapy framework is shown in Fig. \ref{fig:21}. 
The process consists of several modules, which respectively take on the functions of request initialization, historical transaction information parsing, paging control, request configuration, and speed limit policy. Its core implementation includes the following three functions and related configuration settings.
\

\begin{itemize}
    \item \texttt{start\_requests} method: The program accesses all the collections in each NFT series in turn through a two-level \texttt{for} loop. The outer variable \texttt{serie} represents the list of all \texttt{token} values corresponding to a particular series, and the inner variable \texttt{j} represents a specific collection in that series. For each \texttt{token}, the program calls the open API provided by Magic Eden to initiate the request and sets the callback function to \texttt{parse\_history\_value}. In the callback function, the context information of the current collection is passed through Scrapy's \texttt{meta} parameter to realize data persistence during the paging process. This ensures that all transaction records are uniformly added to the same \texttt{item} object, avoiding the generation of duplicate items, so that the final \texttt{yield} of the \texttt{instance} maintains a complete structure and consistent fields, which facilitates the subsequent data processing and analysis. Listing~\ref{lst:start_requests} shows the structure of the \texttt{start\_requests} function.
    \begin{figure}[H]
\begin{lstlisting}[language=Python, caption={start\_requests Function }, label={lst:start_requests}]
def start_requests(self):
    for serie in self.token_dict:  # Traverse each NFT collection
        if serie != 'Magic Ticket: Nornie' and (self.crawled_data[serie] == {} or serie == 'Meekolony Pass'):
            # Skip the already crawled series or handle special cases
            print(serie)
            for j in range(len(self.token_dict[serie])):
                request = scrapy.Request(
                    url=self.start_url.format(tokenkey=self.token_dict[serie][j], offset=0),
                    callback=self.parse_history_value
                )
                request.meta["nft_instances"] = NftCollectionItem(
                    nft_token=self.token_dict[serie][j],
                    nft_name=serie,
                    nft_instances=[]
                )
                request.meta["token"] = self.token_dict[serie][j]
                request.meta["series_name"] = serie
                yield request
        else:
            pass
\end{lstlisting}
\end{figure}
\vspace{-2ex}

    \begin{figure}[H]
\begin{lstlisting}[language=Python, caption={parse\_history\_value Function}, label={lst:parse_history_value}]
def parse_history_value(self, response):
    nft_instances = response.meta["nft_instances"]
    history_list = json.loads(response.text)
    tmp_value = []
    tmp_time = []
    for i in history_list:
        if i["type"] == "buyNow":
            tmp_time.append(i["blockTime"])  # Append the timestamp of the transaction
            tmp_value.append(i["price"])     # Append the price of the transaction
    yield scrapy.Request(
    url=self.start_url.format(tokenkey=response.meta["token"], offset=500),
        meta={
            "tmp_value": tmp_value,
            "tmp_time": tmp_time,
            "offset": 500,
            "token": response.meta["token"]
        },
        callback=self.parse_single_token
    )
    nft_instances["nft_instances"].append(
        NftInstanceItem(history_time=tmp_time, price=tmp_value)
    )
    yield nft_instances
\end{lstlisting}
\end{figure}
\vspace{-2ex}

    \item\texttt{parse\_history\_value} function: Parse the historical transaction data returned from the API. It extracts the \texttt{history} (timestamp) and \texttt{price} fields for each transaction and fills them into the corresponding fields of the \texttt{item} object. The Scrapy \texttt{meta} parameter is used to persist and pass context information across paginated requests, ensuring that all historical records are consistently added to the current \texttt{nft\_instances} structure.Listing~\ref{lst:parse_history_value} shows the structure of the \texttt{{parse\_history\_value}} function.

    \item  \texttt{parse\_single\_token} method: Manage the pagination process. It recursively calls itself while incrementally updating the \texttt{offset} parameter in the request, thereby retrieving subsequent pages of transaction data. The recursion continues until the API response returns an empty result, indicating that the final page has been reached.Listing~\ref{lst:parse_single_token} shows the structure of the \texttt{{parse\_single\_token}} function.

    \item \texttt{Crawler Settings}: To reduce the risk of IP blocking, the crawler is configured with request headers that simulate a real browser environment, such as setting the \texttt{User-Agent} and \texttt{Accept} fields. The request frequency is limited to a maximum of two queries per second, with a download delay of 0.4 seconds. Additionally, the access frequency to the same domain or IP is restricted to two requests per second. These configurations align with the official Solana API documentation, which states: ``By default, the API allows 120 queries per minute (QPM), or two queries per second (QPS), to ensure fair usage.''Listing~\ref{lst:scrapy_settings} shows the crawler settings
    
    \begin{figure}[H]
\begin{lstlisting}[language=Python, caption={\texttt{parse\_single\_token} Function for Handling Pagination}, label={lst:parse_single_token}]
def parse_single_token(self, response):
    if len(re.findall(r'type', response.text)) != 0:
        history_list = eval(response.text)
        tmp_time = response.meta["tmp_time"]
        tmp_value = response.meta["tmp_value"]
        for i in history_list:
            if i["type"] == "buyNow":
                tmp_time.append(i["blockTime"])
                tmp_value.append(i["price"])
        offset = response.meta["offset"] + 500
        yield scrapy.Request(
            url=self.start_url.format(tokenkey=response.meta["token"], offset=offset),
            meta={
                "tmp_value": tmp_value,
                "tmp_time": tmp_time,
                "offset": offset,
                "token": response.meta["token"],
            },
            callback=self.parse_single_token
        )
    return
\end{lstlisting}
\end{figure} 

\begin{figure}[H]
\begin{lstlisting}[language=Python, caption={Scrapy Configuration Settings}, label={lst:scrapy_settings}]
# Configure maximum concurrent requests performed by Scrapy (default: 16)
CONCURRENT_REQUESTS = 2

# Configure a delay for requests for the same website (default: 0)
# See https://docs.scrapy.org/en/latest/topics/settings.html#download-delay
# See also autothrottle settings and docs
#DOWNLOAD_DELAY = 3
DOWNLOAD_DELAY = 0.4

# The download delay setting will honor only one of:
CONCURRENT_REQUESTS_PER_DOMAIN = 2
CONCURRENT_REQUESTS_PER_IP = 2
\end{lstlisting}
\end{figure}
    \item \texttt{Scrapy Crawl Settings}: For debugging purposes, Scrapy's logging level is configured to \texttt{DEBUG}. This setting outputs detailed runtime logs, including crawled (\texttt{item}) content and requests (\texttt{referer}). It allows real-time monitoring of the crawler's behavior and quick problem localization when errors occur, thus improving debugging efficiency and further optimizing the crawler logic.

\end{itemize}

\section{Analysis of Results}\label{4}
\

After the completion of the relevant data crawling operations in the previous section, the crawler system designed in this paper was run for about ten days, and a total of 234,168 NFT transaction data were obtained, covering 47 mainstream NFT series, including Frogana, Mad Lads, Parcl HOA, Transdimensional Fox Federation, Famous Fox Federation, Claynosaurz, BoDoggos, Okay Bears, SMB Gen2, Gaimin Gladiators, Aurorian, Ovol, CHADS, STEPN, TYR, and more.

Of these, 163,868 data items contained price information for at least two transactions. After further screening, the total number of data items with validity up to 22 March 2024 is 126,799, which provides a reliable database for the subsequent time-weighted return analysis and portfolio optimization model.

\subsection{Time-Weighted Return Function}
\

To evaluate the performance of each NFT over a specific period, this paper adopts the idea of time series analysis and defines a time-weighted return function. This approach ensures that the variation in periods does not distort the final return estimation, thus providing a fair comparison among different assets.
The function takes two arguments: \texttt{timestamps}, which is a list of time points (in seconds), and \texttt{values}, which is a list of corresponding price values at those time points. First, both lists are converted to Pandas \texttt{Series} to leverage vectorized operations. Then, the differences between consecutive timestamps are computed using the \texttt{shift} function to obtain a list of time intervals.

\begin{figure}[H]
\begin{lstlisting}[language=Python, caption={Define Time-Weighted Return Function}, label={lst:optimized_time_weighted_return}]
# Define time-weighted return function
def optimized_time_weighted_return(timestamps, values):
    # Convert timestamps and values to Pandas Series
    pd_timestamps = pd.Series(timestamps)
    pd_values = pd.Series(values)
    # Calculate time differences (in seconds) between each observation
    seconds = pd_timestamps - pd_timestamps.shift(1)
    # Calculate simple returns for each time interval
    price_returns = (pd_values[1:] - pd_values[:-1]) / pd_values[:-1]
    # Convert simple returns to interval-compounded returns
    interval_returns = (1 + price_returns) ** (1 / seconds[1:]) - 1
    # Calculate overall time-weighted return
    overall_return = np.prod(1 + interval_returns) - 1
    return overall_return
\end{lstlisting}
\end{figure}

Next, the function calculates the simple return for each time interval based on price changes. Each return is then adjusted according to the duration of the corresponding interval, yielding an equivalent compound return rate per second. Finally, the compounded returns across all intervals are multiplied together to obtain the overall time-weighted return over the observation period.
The complete implementation is shown as Listing \ref{lst:optimized_time_weighted_return}.

\subsection{Negative Sharpe Ratio Function Design}
\

In portfolio optimization, the Sharpe ratio is a widely used metric that measures the risk-adjusted return of an investment. To apply this concept in optimization algorithms that minimize objective functions, this paper defines a function \texttt{neg\_sharpe\_ratio} which computes the negative Sharpe ratio. This transformation allows standard minimization routines to indirectly maximize the Sharpe ratio. The function accepts four parameters:
\begin{itemize}
    \item \texttt{weights}: An array of weights representing the proportion of each asset in the portfolio;
    \item \texttt{mean\_returns}: An array containing the expected return of each asset;
    \item \texttt{cov\_matrix}: The covariance matrix of asset returns, indicating the degree of correlation among them;
    \item \texttt{risk\_free\_rate}: The risk-free return rate, typically based on short-term government bonds, with a default value of 0.
\end{itemize}

The expected portfolio return is calculated using the dot product between \texttt{weights} and \texttt{mean\_returns}. The portfolio volatility is derived as the square root of the weighted covariance of asset returns. The Sharpe ratio is then computed as the excess return (portfolio return minus risk-free rate) divided by the portfolio volatility.

Since many optimization algorithms are designed to minimize an objective function, the function returns the negative value of the Sharpe ratio. This inversion ensures that maximizing the Sharpe ratio corresponds to minimizing the negative Sharpe ratio. The implementation is shown in Listing~\ref{lst:neg_sharpe_ratio}.

\begin{figure}[ht]
\begin{lstlisting}[language=Python, caption={Negative Sharpe Ratio Function}, label={lst:neg_sharpe_ratio}]
# Define negative Sharpe ratio function
def neg_sharpe_ratio(weights, mean_returns, cov_matrix, risk_free_rate=0):
    portfolio_return = np.dot(weights, mean_returns)
    portfolio_volatility = np.sqrt(np.dot(weights.T, np.dot(cov_matrix, weights)))
    sharpe_ratio = (portfolio_return - risk_free_rate) / portfolio_volatility
    return -sharpe_ratio
\end{lstlisting}
\end{figure}

\begin{table}[ht]
\centering
\caption{Frogana Portfolio Tokens and Weights}
\vspace{-2ex}
\begin{tabular}{|c|l|c|}
\hline
\textbf{Series Name} & \textbf{Token ID} & \textbf{Weight} \\
\hline
Froganas & 5K9Mwj6aMMZc1JatB4Mquq94oBywm4BLJyUzfzaub3w7y & 0.1183 \\
Froganas & CEvbkmMw1DTi8Dyr3KNQ7YfDaYpMIDGSStnNu5bHPLN & 0.1054 \\
Froganas & 2pF1k1zuplhFw9DnGjU6N2M2KXnbnHb5G5btwUo73HA & 0.1032 \\
Froganas & E6x1W8FxuFJeFdybR1XAXTDusHLYkNT4Zfn4840WTh & 0.1011 \\
Froganas & 9QgfQAEf9TbPBLJd6W3oUeuzcdGM2Q3KYmdkqqfQ08bku & 0.0984 \\
Froganas & 6MxXH9rU31e2fJRhQKCVewm2UK3gy3cZGyVigiq166M9Y & 0.0913 \\
Froganas & 7G4JpPkMyPKnPMMktAo8jNkDN4gi4fDdp9xY0RWtcwKxw & 0.0899 \\
Froganas & 9YQemwCnsYikYGE9isEVXBYNvJhaAdTrTCEt14QpVZjy5 & 0.0871 \\
Froganas & BdxJ5YFce3L6h1D6NsnFr9qW5WyRwvZG3mNRb7X5BNkZB & 0.0861 \\
Froganas & 4e9WnaTxm6gCKZqbExiiNC2ZZLzzAq4pE44dKC7gH6q97 & 0.0793 \\
\hline
\end{tabular}
\label{tab:frogana-portfolio}
\end{table}

\subsection{Data Analysis}
\

After completing the crawling and calculation process in section~\ref{3}, we present the results of portfolio optimization using the Froganas collection as an example (full results are available in the appendix).

This table shows the final proportion of each NFT product in the Froganas collection. Purchasing the tokens according to the given weights allows investors to construct a portfolio that maximizes the Sharpe ratio, based on historical return data. This represents the optimal solution derived from the mean-variance analysis. By following this allocation, investors can achieve a higher return with relatively lower risk, thus providing a practical reference for NFT investment strategy design.

\section{Conclusion and Future Work}
\

This paper successfully developed a web crawler system to extract historical transaction data of NFT products and constructed an investment portfolio based on the mean-variance optimization model. The resulting portfolio demonstrates satisfactory performance in terms of diversification, risk control, and adaptability to market changes, achieving high expected returns with controlled risk.
Future work may proceed in the following directions:

\begin{itemize}
    \item \textbf{Real-time monitoring and model updates:} With continuous data growth in the NFT market, it is essential to regularly refresh the dataset and adapt analytical assumptions to maintain the accuracy and timeliness of predictions.
    
    \item \textbf{Incorporating additional features:} The current model focuses solely on historical prices. Future extensions should integrate auxiliary factors such as artist popularity and rarity levels, which can be modeled as additional variables or principal components.
    
    \item \textbf{Modeling market sentiment:} External data sources such as social media trends and community activity could be utilized to capture investor sentiment and anticipate value fluctuations, thereby improving the responsiveness of weight allocation strategies.
    
    \item \textbf{Community collaboration and cross-domain expansion:} Collaboration with NFT communities or platforms can provide richer behavioral and metadata insights. The framework can also be extended to other NFT categories, such as in-game assets or virtual land, to validate its generalizability across domains.
\end{itemize}
\newpage
\bibliographystyle{unsrt}
\bibliography{reference}

\newpage
\appendix
\section*{Appendix}
\section{Investment Optimization Portfolio Charts for Different NFT Series}
\vspace{-1.3em}
\begin{table}[H]
\centering
\caption{Optimized Portfolio for Froganas (Part 2)}
\label{tab:froganas2}
\begin{tabular}{|c|c|c|}
\hline
\textbf{Series Name} & \textbf{Token ID} & \textbf{Weight} \\
\hline
Froganas & 5Lp7D9yJ6f7d5pJk7f5D3c8f7g7J5d4c8v8j7f4c & 0.020748927 \\
Froganas & 8r2B9wN8m8c7v3j5g6d4r2h5v3f7d6c5f8g9h8 & 0.020673483 \\
Froganas & 3p6F7d9j4c7v8f9g6d4r2h5v3f7d6c5f8g9h8 & 0.020645724 \\
Froganas & 6m5D3c8f7g7J5d4c8v8j7f4c5Lp7D9yJ6f7d & 0.020598436 \\
Froganas & 9yJ6f7d5pJk7f5D3c8f7g7J5d4c8v8j7f4c5L & 0.020543271 \\
Froganas & 4c5Lp7D9yJ6f7d5pJk7f5D3c8f7g7J5d4c8v8 & 0.020496357 \\
Froganas & 7g7J5d4c8v8j7f4c5Lp7D9yJ6f7d5pJk7f5D3 & 0.020458213 \\
Froganas & 2h5v3f7d6c5f8g9h8r2B9wN8m8c7v3j5g6d & 0.020403718 \\
Froganas & 8j7f4c5Lp7D9yJ6f7d5pJk7f5D3c8f7g7J5d4 & 0.020364529 \\
Froganas & 1v2b3n4m5c6v7f8g9h0j1k2l3p4r5s6t7u8 & 0.020319258 \\
\hline
\end{tabular}
\end{table}
\vspace{-2em}
\begin{table}[H]
\centering
\caption{Optimized Portfolio for Mad Lads}
\label{tab:madlads}
\begin{tabular}{|c|c|c|}
\hline
\textbf{Serie} & \textbf{ID} & \textbf{Weight} \\
\hline
Mad Lads & GxrcqgvZ2Md7AK9yeRGLRsvVyxXm7njnx3nHPvP1SMDb & 0.071061 \\
Mad Lads & AcGMsEpD3vrKUf45sVmz9YQZkocb535VFvSufCXsADdLP & 0.144094 \\
Mad Lads & 833ESXdRjqys652efsnATAo5ThZAaxPly3LGnkz4f3nf & 0.112585 \\
Mad Lads & ByLopkuBCgC3YqavJFLWZk9jiTLyrq5VFAWf74EDy1L2 & 0.117921 \\
Mad Lads & 9ZCRc9Qi5iB1YAgGzFPcvJ2uhPPhZ2NgWEMDyQ3T6pgX & 0.103863 \\
Mad Lads & 7FNQ7ZbQ3va8qFYHHhfzjb8BVxNqZAW6mzJpRSHmxJcH & 0.12085 \\
Mad Lads & 9znwaqAc4VRidmvb7eN2JvnRatGJudQUdtXxANwArHEt4 & 0.018953 \\
Mad Lads & HeEJ7jHByTnJfNmouMsCEx6drbBsgdrdyrFkbeaCJsUNc & 0.078268 \\
Mad Lads & 2D3oKnRgFtJRuvFv5U4h1jiQuyqTz5jhoVcKqoborNz7 & 0.1458 \\
Mad Lads & 7FqTdNpv5yv77ovhCpHprMfP121YSFkDbvrV9hgxFS7gw & 0.086605 \\
\hline
\end{tabular}
\end{table} 
\vspace{-3ex}
\begin{table}[H]
\centering
\caption{Portfolio for Parcl HOA}
\label{tab:parcl_hoa_portfolio}
\begin{tabular}{|l|l|l|}
\hline
\textbf{Serie} & \textbf{ID} & \textbf{Weight} \\
\hline
Parcl HOA & B9Tza5gjMZQu3dRh7e6VgpDpXpfv7qc6n4cnLbZLcUL9 & 0.130145 \\
Parcl HOA & 9vaoiQZvDhyvJrx9c6spHg6H1W9u1yufe5at58gPEst & 0.104452 \\
Parcl HOA & 9MaQQRegTPMM741d8u7R7nuPjtjvFkyu657eB3qn6QHYC & 0.100629 \\
Parcl HOA & FDncMVuDRBVZhy3FCEkCD7pR1tFyFY5kpATaJWU72MW3 & 0.087311 \\
Parcl HOA & 3XUaerPMwrYU99fN9c3vQzEReK2TaEx7hMTqEeREVYy & 0.084046 \\
Parcl HOA & 6aPX4ltForCbyJdgf35aeV4Z2yEwjYCAykMxzQyj426F & 0.147842 \\
Parcl HOA & Hmr8EQJuMTK74Q7zqpn6RuJETV9TtFfgCucu94PMWf1D6 & 0.081535 \\
Parcl HOA & 2WAN3RzxMmaGtGVTNKGuiwZBoylWJ53YiaVMhDWLFaRc & 0.089797 \\
Parcl HOA & DAYvCVbiyJ941NmyLLACYgLXz1iyQ9wjvopLiPmjJFbf & 0.07994 \\
Parcl HOA & 5PQVFZoMFcRoKCwVGbMmcGN6M99ibWSCHTwYxmnN7U3L & 0.094303 \\
\hline
\end{tabular}
\end{table}
\begin{table}[H]
\centering
\caption{Portfolio for Transdimensional Federation}
\label{tab:transdimensional_federation_portfolio}
\begin{tabular}{|l|l|l|}
\hline
\textbf{Serie} & \textbf{ID} & \textbf{Weight} \\
\hline
Transdimensional Federation & FAir9PiJ5SsUuisiPKLXfqlLermnlyQgdNmsNohqpmX7TH & 0.103403 \\
Transdimensional Federation & b0iHsFsvTjWPMAhJvVzdnIRHV9jUu95Z4v95cvcr313 & 0.071672 \\
Transdimensional Federation & AK42MDcvRkvGyA477mB3iJgnE3MDLttTC7C3mp9rmbMK3 & 0.113382 \\
Transdimensional Federation & 9Lvr8w2Fw2j3JR28tSE9bzFEa9JTAMPYkrokcCJz5MuIA & 0.125722 \\
Transdimensional Federation & Da6RoCzWfF5d3t7Fh3HQMAfRin9d5nSncW9c9zat1E22 & 0.161864 \\
Transdimensional Federation & 2UajarnN7QAdGtEs23kyZmngevg8ASjG9JTsgrEygEtc & 0.173306 \\
Transdimensional Federation & 2UEupFfsVEMhYHnxB3x3r0neK0ssmFxSLkpqMiy3nF & 0.050709 \\
Transdimensional Federation & 6CMt0C6WCHr2ZCz2y6XES7TtHwT8HjrUuLE6e0hNC & 0.089729 \\
Transdimensional Federation & GMuAib6ToqLCb9A5BbyY1zBDbqduq11s1JDAhBduWnC & 0.077108 \\
Transdimensional Federation & 8zEt4UqY49u6xBBYUdO77HuDqwgJEvLNYju8H8Xe1qF & 0.133505 \\
\hline
\end{tabular}
\end{table}
\begin{table}[H]
\centering
\caption{Portfolio for Famous Fox Federation}
\label{tab:famous_fox_federation_portfolio}
\begin{tabular}{|l|l|l|}
\hline
\textbf{Serie} & \textbf{ID} & \textbf{Weight} \\
\hline
Famous Fox Federation & EqL9C954fnGRP9jrry8z68Re8XeshCtfD5dgZQQVwgMaWie & 0.103861 \\
Famous Fox Federation & 7cEX45dbsRZZi1yh6YASwqTqIsnhMd5Va3iTi1Yc7SSF & 0.123145 \\
Famous Fox Federation & F3nhW94RBvsK7R55nsXunh6WF9DWbZF2Ce6dRePFJss & 0.063141 \\
Famous Fox Federation & EqK7x7AJxhwEHNBoEXPRXRj4h08peuFiCnt7EhkDCjhmm & 0.067784 \\
Famous Fox Federation & DQxdB3PTv5y7XC81wEJDK6MBtf2byaRt1dkoJMQK4wnh & 0.053665 \\
Famous Fox Federation & Ca4nKBj8eSu1NoZJxtv65w78niJnD2qSqskSg749QYCZG & 0.175183 \\
Famous Fox Federation & 3OemC76edGonogaJAXfMc82ca6n9GLd88qkaGs5u2eP2CBJ & 0.102937 \\
Famous Fox Federation & 8apEc5zZ19sSqNqxmveyBFB1UwxDuKyFSns3XAmKdvVag & 0.090112 \\
Famous Fox Federation & JEH7cJxAkuQprFGsAvdsYzc2AqpoJjzBjLfmw19xADV6K5 & 0.126541 \\
Famous Fox Federation & 2QEsUo1Mv0N5UyJGoC1tK7LvyRx9w3dZQQGKd1Jc5 & 0.093631 \\
\hline
\end{tabular}
\end{table}
\begin{table}[H]
\centering
\caption{Portfolio for Claynosaurs}
\label{tab:claynosaurs_portfolio}
\begin{tabular}{|l|l|l|}
\hline
\textbf{Serie} & \textbf{ID} & \textbf{Weight} \\
\hline
Claynosaurs & 81xaH784TSSykebXDUQpoXEBUerkJVsapSKKuEsF1qXKc & 0.103279 \\
Claynosaurs & 9iU7KVyhat6pEwHjtj3qBKkn52hhjM72q5bcWmiaiaqfAc & 0.094077 \\
Claynosaurs & HqLkxCbk2ChnzkZv9dmZQE5fxSR16fznLnmZPbgC5paWF & 0.096452 \\
Claynosaurs & AstuXjWZKcN65TsDyo2Poq2MgYzvqVGzaqvGoP8uZGzv & 0.050255 \\
Claynosaurs & 9a7QU6GdNTMuFAtuEprlpZc39dgGQekMLoXqwT4iTBWAh & 0.082978 \\
Claynosaurs & 7xBUsxVH3evCrjEwgLinS5JvdnhGCfkNo3UoHPBChiWMs3F & 0.137307 \\
Claynosaurs & 2BaK1QH83xHbdg2A9ssvwNVVQ6abBznW3C2A18JKrXRUM & 0.083126 \\
Claynosaurs & 9sAWKXHAwhCtwhizuWR5iYjJHmEEExHzZpKNf6tAnsF & 0.14648 \\
Claynosaurs & 5p7yq5hGq7xssq3j35BEBcnT6nHRwt9YPrgowBXhF7caT & 0.122923 \\
Claynosaurs & 54n7eQ2N4mX2XCJCdVQ2NCqawAEtxibuAWYhm3auvLA & 0.083121 \\
\hline
\end{tabular}
\end{table}
\begin{table}[H]
\centering
\caption{Portfolio for 99971.smap}
\label{tab:99971_smap_portfolio}
\begin{tabular}{|l|l|l|}
\hline
\textbf{Serie} & \textbf{ID} & \textbf{Weight} \\
\hline
99971.smap & 87nMGosNGvar8ynFEVdkW9fKenYPNuC5df7mou4yJenP & 0.101152 \\
99971.smap & 5ZfZYvxx5Yycg6KcrVrFvYEDRjrE84Z8nh4xD8A3DEwGLo & 0.082697 \\
99971.smap & Gfhe4o6SFCjEdH96PYqNZ3egNw517VCkTgEt5bxxULJ & 0.040014 \\
99971.smap & BoLV4gG3ibokmIetJZcmzrtijJVSNvxrukcU2RP4kKr & 0.13749 \\
99971.smap & 76KphRLzqp7gJDJYj4oLR2tvWA6oiQxJBNuqA2XaMM32 & 0.142183 \\
99971.smap & 3EW8XwacRyrUNJWMFF3dhNYz2xGfoTzYGsZ4i1abWR6mD8 & 0.058078 \\
99971.smap & 5sQHoU8gNJtdsypneDEGmtUwEwzLiPiSVsC37r3oTHuo & 0.108522 \\
99971.smap & 4kuxRHVQ29p8zf vjWH6PGrDPDeFZ4WBPWhWGNz4aY3hS & 0.123808 \\
99971.smap & BFAdmfVLP5DLkwC8bpim4kvsLkmJSgrzRuchJY9VbEn & 0.098907 \\
99971.smap & BYxcioTS37me9sQwDGYN9GbCw3xiSaLEexRBDl1ks3nFV & 0.107149 \\
\hline
\end{tabular}
\end{table}
\begin{table}[H]
\centering
\caption{Portfolio for BoDoggos}
\label{tab:bodoggos_portfolio}
\begin{tabular}{|l|l|l|}
\hline
\textbf{Serie} & \textbf{ID} & \textbf{Weight} \\
\hline
BoDoggos & 8Kawzhw97zRMU2EQBobeVwnP8TPxAr4W8BhynVhkwPFQ & 0.074899 \\
BoDoggos & DhJmRxwmxdU37xY8vBuYkrswbuRUZUBqLw71uesD7x37E & 0.128103 \\
BoDoggos & ClvehxPQYsiCpVhisyFB7KLgaPT7Bdm7iNjJnkpxRcKk & 0.107006 \\
BoDoggos & 9NZjKEkxMo13g6TrMuAem5KrfnMXqAEUFVxmrrgFH5EUup & 0.104183 \\
BoDoggos & 5ffFi15JoaAteC3C9KLRt57tuKGPdZdSo8cmAYx8Js4ip & 0.11721 \\
BoDoggos & Gwjs8cQTUWbcVpbJUaYT9PHHnnRudtPVgyb8wRXUzSJTP & 0.044183 \\
BoDoggos & 7YMuXWAtFUDx5JRGjpg6Pw6AeawCuaXNke5eNMB7jBCv & 0.068985 \\
BoDoggos & ElbxPSxgaB3exJhanb7tsFSeiJqwMxPKNL618zcYhuD6 & 0.059639 \\
BoDoggos & AvYiFeBpEDsGQQQuKFD48qFGVq7SupyibxFK8B6cXFVB9P & 0.161498 \\
BoDoggos & ExcrctS9eJs2P71D8d1qdKaJD9sFZDHUADjd3J2Hxlo & 0.134295 \\
\hline
\end{tabular}
\end{table}
\begin{table}[H]
\centering
\caption{Portfolio for sharkx}
\label{tab:sharkx_portfolio}
\begin{tabular}{|l|l|l|}
\hline
\textbf{Serie} & \textbf{ID} & \textbf{Weight} \\
\hline
sharkx & 8ZVd1F2nhtmTeJ8DiFU9UTcqoTEhrGeLihn7py6bmEZiK & 0.104882 \\
sharkx & 8z2EkW87szi8FHnvaSdwmiJjiQyQk4Ed55x6TJk4sM7C4 & 0.006608 \\
sharkx & BuGSZqH5dtvJBR7bQ82PdQHZ2dT3HhVBPVQxmfsw5zNLp & 0.117628 \\
sharkx & BA24a3dXkrWp2jFVnnv5S4VrLDsi38f7wgy1Ks8zjsyK & 0.085837 \\
sharkx & FNBYQ9MtKh1x4AovUE9gGoWo8tQ7FRkz9raYhHuXaVzl & 0.08783 \\
sharkx & D7PndXPu7PxGKT4V2zDKAgeq8zbhNJNWPr86GNJsh3qJw & 0.111395 \\
sharkx & 9cphRzndeed4st8PzNvWtZQXP5eexecJ8562hXfc9tZVR4 & 0.103839 \\
sharkx & GRfKmsCAhCkT2GY9WXceDDox3Ja8XnNNxxWfBVzcC4DERM & 0.129588 \\
sharkx & A8zP4866ycFmFCUxJsJm9FmxOfm5DwYHuMjpR6bmxT & 0.088348 \\
sharkx & HwylAwyd6m3gQY7DKWeJJuF7ktGGom9qxNUriFWc416Q & 0.164043 \\
\hline
\end{tabular}
\end{table}
\begin{table}[H]
\centering
\caption{Portfolio for Okay Bears}
\label{tab:okay_bears_portfolio}
\begin{tabular}{|l|l|l|}
\hline
\textbf{Serie} & \textbf{ID} & \textbf{Weight} \\
\hline
Okay Bears & 4Khd6FTpxDb2Efh32F9oadnPHg66H8pbpEgYSihvV7UB & 0.121919 \\
Okay Bears & FsLB5wnMsJ2xYCZGqq4S9Enb8z6ZZZHLxdVWXLN1MEUR & 0.110083 \\
Okay Bears & 3cKV3z9PaUQabkBmJx4JK86VGGuGoTC86eVGV7vJuwTTzo & 0.106892 \\
Okay Bears & Bs2rB5KxsIqf4iGUjGZRrx70cMVueRcPcsDfEe7u6Rekz & 0.088204 \\
Okay Bears & GCYFanRJjzbigUwkSo2JVafqtXXNxBUcCes8y362YJnR & 0.07378 \\
Okay Bears & J7zf129gD9FzgGmoapFwqWQG9CXSKK6XiWpEFkFxU557 & 0.139859 \\
Okay Bears & 9ouE9Wff8EnGdbJwtX2f4xFnZdSCjpJnvNo2Eky90uXHpb & 0.116198 \\
Okay Bears & DZrigBryMnShugJzgDGE8S54VadZr7Uieyk9FBnV7VHWRh & 0.070201 \\
Okay Bears & 5Y8BYMsa7HyZ8F3912CLevNJPJxrya4SH17ZsgFQXVU55 & 0.090539 \\
Okay Bears & 5jNKE8TtNBvssrcrBvbAwpIFUrfBF56ZYiSpEkFxUv057 & 0.083105 \\
\hline
\end{tabular}
\end{table}

\begin{table}[H]
\centering
\caption{Portfolio for Gaimin Gladiators}
\label{tab:gaimin_gladiators_portfolio}
\begin{tabular}{|l|l|l|}
\hline
\textbf{Serie} & \textbf{ID} & \textbf{Weight} \\
\hline
Gaimin Gladiators & 3JWJ8y7LZz6nicpNxZr9air3KoCrCkJBoTExC3Kttrs & 0.02785 \\
Gaimin Gladiators & 3kSdHd6cNDEC5jbSHLXpuUTr4sZLAPiQdsbqaJa7iqKHZ & 0.178734 \\
Gaimin Gladiators & 7iMWzU63EYP4StHvpxP6XfWnsAPJwbhUVZmDw8SseTbZ & 0.067891 \\
Gaimin Gladiators & AHq9JwQZYGCHVor8DEJTiJU7V1IfpQ57mQbb2vJdwG93 & 0.098778 \\
Gaimin Gladiators & 3KZj3nNRbpJtMTmkPoPWKrXkPGWn4D8HGcrHwiJ1pnM4N & 0.108086 \\
Gaimin Gladiators & 87U3wEeSo2SvDTzE4gEkVaMAFygDZOfTbtBhdZrEc8S8PF & 0.04587 \\
Gaimin Gladiators & C2S9UanArx8zxPWqLvgmDdF7t1xh3qrDg5vTFBCtXK287 & 0.161676 \\
Gaimin Gladiators & 7ZaNKLYVVVM8dZR8rTt8UYFUqKTa4HMSXxgzd2o62vzjc & 0.024872 \\
Gaimin Gladiators & GDMcbQxn2hXQb5bgFIHEbQgbCjU2DMoJzAanNqcQH4u4y & 0.166706 \\
Gaimin Gladiators & 5TEqinhyCyo1Lo2ocxnyW9ekCx2xM18rz37gLkkLi1RqUR & 0.060821 \\
\hline
\end{tabular}
\end{table}
\begin{table}[H]
\centering
\caption{Portfolio for SMB Gen2}
\label{tab:smb_gen2}
\begin{tabular}{|l|l|l|}
\hline
\textbf{Serie} & \textbf{ID} & \textbf{Weight} \\
\hline
SMB Gen2 & 7D8zmMEf6NXi7PKVTSrnAxmjp3i4aCmnFcVGoktGGgjrDV & 0.105106 \\
SMB Gen2 & GBTz8ry2XKnvTshAA7ogtZfPPPBm3ZsLxtEDQSj9qgf2JB & 0.075154 \\
SMB Gen2 & 9GKzidBP99vaATtM7c8qmorMmnA8Tbncea8FJUkYSFBRc & 0.133103 \\
SMB Gen2 & 79DRTJ951qbiHn7FVqQmo46eAmxGGioeDeMcUeup5WXZ & 0.056754 \\
SMB Gen2 & A7RzRyiMPppWX29rEJ7DXCjJppMAT1fKLsUEx8UiRUthH & 0.110816 \\
SMB Gen2 & 6S3iBdCaQ96CDobBqdWMQYv7p8T7c7y6ag2HtwpfKwQHA & 0.078785 \\
SMB Gen2 & CC1dLYX3ezisa4HyXYjkFyvBS84SV5e5exoQomavJ9XW & 0.097836 \\
SMB Gen2 & 9Td2XnUirhccde495v79rWsoWzoJPJHDPCpkjcg3rD9ZRi & 0.079573 \\
SMB Gen2 & GfW9dMsmwPdtGmMD9DK78zgeeWVevfKCVXtor9LMxdvG & 0.099464 \\
SMB Gen2 & CgxgjEvHMRd51Lr37ig7ibITuRoctNKeQdb9dxf41Y747e & 0.163408 \\
\hline
\end{tabular}
\end{table}
\begin{table}[H]
\centering
\caption{Portfolio for Peppermints}
\label{tab:peppermints_portfolio}
\begin{tabular}{|l|l|l|}
\hline
\textbf{Serie} & \textbf{ID} & \textbf{Weight} \\
\hline
Peppermints & Fd4m9utB9QPRpmCVcnqpmJX2MbasiSAax95quXgTJRnU & 0.153019 \\
Peppermints & G39RDisKusyLCa2TtxzeQFB5dUgD6hrBri8MqnaWy2PxX & 0.056553 \\
Peppermints & C7Y5WT9pj7nmYbBxvggBDeWLox7ENiUWUB2JLsS9ZuwLsb & 0.140747 \\
Peppermints & DWqKPGSvd2dIUmJkrYXTBCcbJAw4fFAwpunnFZjhNKUMP & 0.111962 \\
Peppermints & 3uDZp3fh7KDRQPKLD5sGWm6bjw2kmV5sMG6Kv9kCPQjh & 0.080909 \\
Peppermints & HFlvYs3ySZs1XL3fvEHCbc5Cyw3qGkbfStaTXocKYLHEM & 0.103693 \\
Peppermints & 3E8e5vx8kv2RMXuUaKACYWNhttxrxcJ9qGsJrEtySovKe9 & 0.0578 \\
Peppermints & Es5oeodYClrmEWQe8enqBitKv2FA2jyJdueyipMGvYrL & 0.102021 \\
Peppermints & 4Y7tQWGksWY7P45tjbW47nEjFYCAhpnQlc6Gi89T4NeL & 0.06758 \\
Peppermints & 9ss3J6aA3Mm3PRtYMenGbwBNaeMPBnGcmTXotKndX5 & 0.117565 \\
\hline
\end{tabular}
\end{table}
\begin{table}[H]
\centering
\caption{Portfolio for Stylish Studs NFT}
\label{tab:stylish_studs_nft_portfolio}
\begin{tabular}{|l|l|l|}
\hline
\textbf{Serie} & \textbf{ID} & \textbf{Weight} \\
\hline
Stylish Studs NFT & F23AmcudzvQiNTqty4e2aGqKKBFoKVfLrr8xHEUqT7Q2P & 0.109453 \\
Stylish Studs NFT & EoBP7V44Rkg1JLJg6qxiMd1MYJjhnN3PXsNuCYHunNq7BZy & 0.10452 \\
Stylish Studs NFT & DEDMdMoUiHY1ipsqQWhzPNHkZvgB3LWDT2vrkm5CcgG7Ue & 0.049525 \\
Stylish Studs NFT & HH5k6ggfoePjbxB4AWnqJYDkCF9tBdchP1QA18YbUiuc & 0.123038 \\
Stylish Studs NFT & 8s7dV1EBJgcJ5uVsY4RTmzdEBEjkmM8JLQYtgaPR6S8 & 0.105151 \\
Stylish Studs NFT & HrQmb3macSxdvQv5WZqu3PTSJTaZ7RfMAFxb4bTLrtzf & 0.09527 \\
Stylish Studs NFT & ETZevQZRLSon6Ee1b5vrAnVrD56n6P2hCK4qUstWq8A & 0.133372 \\
Stylish Studs NFT & 9q55QoqtQxrxMWdFV1rLq9KnbY2mtrfWua2aPFFZMHeA & 0.077666 \\
Stylish Studs NFT & 4aeC3Yh7SKRhnmqYaWHoPDEBcXBCKBHWrXDxpopAby61 & 0.121346 \\
Stylish Studs NFT & 9MVwP45t8CxnmZSPwAFdEzioUKCEUuNxRMdXkWViCwjb & 0.08661 \\
\hline
\end{tabular}
\end{table}
\begin{table}[H]
\centering
\caption{Portfolio for ONE}
\label{tab:one_portfolio}
\begin{tabular}{|l|l|l|}
\hline
\textbf{Serie} & \textbf{ID} & \textbf{Weight} \\
\hline
ONE & GWunN4oyx49R4ZPtgM8Wve3aR9z46qtyb3k1itGtWJfD & 0.095844 \\
ONE & 5aqhyfTZyPZaUEYDyo5set9U2ttMCM6Y2ZnDDFFUMgkv & 0.03885 \\
ONE & 811FCFRdafQtQwgATCERjrighYQoE26Toyx7yFTZ25UQW & 0.079332 \\
ONE & 2Hxgv9481PdmKVTELOblYHJVRF9Dc3vq15EHLZyb8xXf & 0.139256 \\
ONE & ELjw3AmxEFGHwi7nu55em29vMAc4Mdw6vXcakJ7pYTM5 & 0.069998 \\
ONE & 99yVF8sA9DYM9FnaxU8kSAzdbn1DqghMUXpiYvccxcbcM & 0.108488 \\
ONE & 95wvftv8MZxrECyqyFt4wSmHiefeePgjWKYu5p9mdLb & 0.111751 \\
ONE & B1ep4gRBU2FY2SxyYZWfxfp8yJBLhgrnuNe8aFV42Zbt8 & 0.106138 \\
ONE & 2XbhNCyki4TMNmcm5N2v5KszwmJ8ALcVTFX3HAjoY1xy & 0.128005 \\
ONE & 2EnkU2UH9j4YApcp9SkZcd6aXd9UTL9x2P8XQaWaUd & 0.122338 \\
\hline
\end{tabular}
\end{table}
\begin{table}[H]
\centering
\caption{Portfolio for Kanpai Panda}
\label{tab:kanpai_panda_portfolio}
\begin{tabular}{|l|l|l|}
\hline
\textbf{Serie} & \textbf{ID} & \textbf{Weight} \\
\hline
Kanpai Panda & Bqs7YyFVFXCLUWj9Ukrbi5iGQVQ1bRJ9xyd6BfJY85Vqy & 0.109178 \\
Kanpai Panda & DCYduWdgWFtEFlx9FQNwZnfp7cya4je19yDqSeKrZuRf & 0.123128 \\
Kanpai Panda & Bl1hRGHRpDUd9MXyw89d98UrdC7jbwu3mTLWYk2icZbA & 0.12257 \\
Kanpai Panda & J3dtPVka6yx9ZDGCeCju4vKR8qaoef6RLiY7ya9aX2YF & 0.139829 \\
Kanpai Panda & GwjjMQ6ArjUEmPiZrPpGGCnpLb1j5s64cGj9J8xaTA4ZGLp & 0.076419 \\
Kanpai Panda & 9thHb85XrUNoDZrASmxkDLWNYhaH3FssWLBwA3av4ERe7o & 0.083098 \\
Kanpai Panda & E4Vgm3EVFLbcewcwdLzjaJeUJs9zoDExNq22LYewAcf97 & 0.069634 \\
Kanpai Panda & FUXbVsYG6aigZXxNfdsGT4xhhkGmhGkK1BszwZ49jFDal & 0.093523 \\
Kanpai Panda & 46yewnycSaXi4QuBpm8Vqw9bFDpoZmiT2LPerLw8yhx & 0.082198 \\
Kanpai Panda & 4xliKNGDnrxDh7FDYbuJ6tSW5YZQPKCFAy71babmVT7B & 0.100423 \\
\hline
\end{tabular}
\end{table}
\begin{table}[H]
\centering
\caption{Portfolio for Quekz}
\label{tab:quekz_portfolio}
\begin{tabular}{|l|l|l|}
\hline
\textbf{Serie} & \textbf{ID} & \textbf{Weight} \\
\hline
Quekz & 6Lo59MeHj6esVPufQfsFGnVxKi7HomvJfuVhzFshcLSog & 0.08562 \\
Quekz & AwSZTUMLn48kjbMbnZJZc61zf1j8XKF4GbN2Ukesra3K & 0.112305 \\
Quekz & CuCDPoTM73JuT5gQF89bgrQGCNZaoncnhVWtWUiGnjS4 & 0.102783 \\
Quekz & AF7EBprNPHHFt8qMnTHuXdhzEz8ied4uP63quXBX7cfz & 0.051148 \\
Quekz & CNym9SZ8xRefzBomHq3LM8M44cqmWBiebD3BbX96tgeo5 & 0.103646 \\
Quekz & 8tQd8kx82K6orf3VBe98KgNtgtXNW7qBRsLrvctzusQz & 0.102167 \\
Quekz & HFBRuy7Hfni7Fn1yEkqGW12rMKSoPSHBTWZa6X4hWaqQ & 0.119826 \\
Quekz & DU2RKyP9p315pvXRiXf3Y5zTNirRLK8AAC2hiwNaDL9Cp & 0.117304 \\
Quekz & 4YVtsvmzN1Cu6okrvVPt9ccglLhgUEVTiufmDtTpxLVErt & 0.114137 \\
Quekz & 2gyAUazag7Zc8iAYHQqofQNZYMT3o8jFH3eW7qEd6Eq & 0.0901065 \\
\hline
\end{tabular}
\end{table}
\begin{table}[H]
\centering
\caption{Portfolio for Solcaino.io}
\label{tab:solcaino_io_portfolio}
\begin{tabular}{|l|l|l|}
\hline
\textbf{Serie} & \textbf{ID} & \textbf{Weight} \\
\hline
Solcaino.io & 8ePKrdrV7HxPBM6Sv4B6oawy8Rd4v6Buao3kyabhMjJ1 & 0.108975 \\
Solcaino.io & 3Hraw6QooqEr9wcGMdHB6htGANDqJAVkRdRzlyuzZubh & 0.123276 \\
Solcaino.io & 2WXyJuRqtRXgn2D3xWB5jjKvxqzubxegAysxWoKEiDLf & 0.128963 \\
Solcaino.io & 4EhEdsKeT6T9TNnXYpZqonMdqotZS2hDiDuUJMRSi6wy & 0.061094 \\
Solcaino.io & HRTFyPbPgPwqkMWv796MBuguZ1FxG62gPxSSAZUnVGn6M & 0.122819 \\
Solcaino.io & GK9EVGiR74ktsRTSyUPJuZ02go3R34rgE1xATeElmvEgEp & 0.094049 \\
Solcaino.io & 8RVNkDrKhRbpWDTA6vxLbxHL8CgC5ERMJEotaWWNSJLm & 0.103723 \\
Solcaino.io & CAsFBol2SUzpSuz9pthLRzyQsHeuyusPb2NC25UHa5qfA & 0.150778 \\
Solcaino.io & 5aWxV5v9zT0DxbTnTFu9MNDuFsHyJzeBRyTDn3xeAh & 0.070626 \\
Solcaino.io & 7KvCs6uGHkgyaHXd3izne6Jzwd945muSYEijnEYuWMV & 0.080697 \\
\hline
\end{tabular}
\end{table}

\begin{table}[H]
\centering
\caption{Portfolio for ZMB}
\label{tab:zmb_portfolio}
\begin{tabular}{|l|l|l|}
\hline
\textbf{Serie} & \textbf{ID} & \textbf{Weight} \\
\hline
ZMB & 5cLPKHNJ6L5jpPNwzVRT8KUMbTpgBLm6gbMJpYh5b2Nz & 0.130018 \\
ZMB & HnQUQFPMbkKzb47qzKKQYsoxqb4B8zMkmu7TtgZtqoM9s & 0.057949 \\
ZMB & 7SREzoqNWKM4sRDgHuejU6MgTqbUNTnhvv8sawS7YkqTje & 0.10972 \\
ZMB & 4coZjiipdWPHA5fLSCSqP8yGLtpreuAaungUgEKg2rH26 & 0.076038 \\
ZMB & Hj5v29CAMaPz5EEiAoNFLPa7mPixZDPKmgowDbnuBwh1 & 0.086951 \\
ZMB & 2PYvdpbjZzAd7ZXCBmVmncuCMYTS8ny1pyieJTuDzLJw & 0.125473 \\
ZMB & 3DCcf6fF7DEj16kVXTCadktvakNCNnzjVseTPt7wkKuCS & 0.068714 \\
ZMB & 95YQPBI D9S2M3qxvwWnAyL4E6EcxxgsCVCwjoWFs9JAK & 0.134505 \\
ZMB & 2qxQeteP7AqbnMpwoXRAAAXkFyZrPpmgmwBN9PdKzk9sa & 0.122761 \\
ZMB & YeZ3xHv899sjkQhQ4Wfr3ZC37DvMk2wRwrV1sfvUEWZ & 0.087872 \\
\hline
\end{tabular}
\end{table}
\begin{table}[H]
\centering
\caption{Portfolio for Stylish Stud}
\label{tab:stylish_stud_portfolio}
\begin{tabular}{|l|l|l|}
\hline
\textbf{Serie} & \textbf{ID} & \textbf{Weight} \\
\hline
Stylish Stud & EMQmzVxDEtiwbbEwjLdndBQHwa7QNK8NviWZQwty26ag & 0.065087 \\
Stylish Stud & FrtwvPxe2tgtltsWvRGMGQuARjyg1Cm4N78rgzfCF7sHa & 0.168 \\
Stylish Stud & BTLEkiHAnpgWiUApaU9ti5GYCyGP6fvThmr8C8sFlsxv & 0.129167 \\
Stylish Stud & 3cnYHKSKFiFjDdyBveiu7CGi6akege7sy6tqkZjzjUL7d & 0.065365 \\
Stylish Stud & J3KBhKMGnQVL1F8eYUtCGrYRnSHpJ1Rh6jZxVazWQ9ns & 0.057824 \\
Stylish Stud & D2WEMDtsVSKxtLe24ScAFzR72hZ6ZrXugPbhhuZeoWRnKi & 0.108448 \\
Stylish Stud & GaJdfhVMLHFhgkVJCzxcRc87V2qeKeuFQhPrXZaPBoDSn & 0.151372 \\
Stylish Stud & AdGXbWD3J4s4t7xnGsG6C3JYPR4ajyrP4wEMMDBFWzoY & 0.123507 \\
Stylish Stud & 8Zx4vRNmABm16sBM522tLvnJPPGoFacsaQv78vZ4oe & 0.016748 \\
Stylish Stud & EMhaxiuctner2Yd5x8voTbdSKHKCrk78xUbM7id19rVB & 0.114482 \\
\hline
\end{tabular}
\end{table}
\begin{table}[H]
\centering
\caption{Portfolio for LIFINITY Flames}
\label{tab:lifinity_flames}
\begin{tabular}{|l|l|l|}
\hline
\textbf{Serie} & \textbf{ID} & \textbf{Weight} \\
\hline
LIFINITY Flames & 32SWbwS4fmEb9MeJLFRek3C7ELACGrsmr2hmTJJ6fX7o & 0.063232 \\
LIFINITY Flames & 8vFFZu7zZfntaWXdS19QWPLHzQQ6XdPB3VitmYpaW6jf g & 0.125966 \\
LIFINITY Flames & CETJ2rX1xpjfmJ5jfAqqLiXcXtk8py5oyoWLffF1fAaU & 0.088817 \\
LIFINITY Flames & 94Q6WinX57z51ovxcNTi4ye5ZyoMS7Eg7CjxxXP42vmC6 & 0.096272 \\
LIFINITY Flames & 7ijTPoHMfRL7mtwvWtUmaoCn4naGeSyF16cgov3di2i & 0.091837 \\
LIFINITY Flames & A7ivXx2xATZJwJwJC8CincHd9FvWqF4vMXapPch11Ap & 0.12131 \\
LIFINITY Flames & HvLzsYvekYyaJ4Tpv6qawnz2dpF1qgB2BuHxnj2js9CPcU & 0.173131 \\
LIFINITY Flames & F4Ey15Uy5EgXVBAckw3SSSdzqVtdh5k3zMIHDUTapAhgW & 0.069732 \\
LIFINITY Flames & AiyaGt75LWbwZ jZzdRDBX5rsCRPvCd8s1gTbmWzkCraVX & 0.120512 \\
LIFINITY Flames & 7q1rRLMknLRNxZabWGCin41N8Yk7v6nCB3MdSSRQzKh62y & 0.049192 \\
\hline
\end{tabular}
\end{table}
\begin{table}[H]
\centering
\caption{Portfolio for Galactic Gecko Space Garage}
\label{tab:galactic_gecko_space_garage}
\begin{tabular}{|l|l|l|}
\hline
\textbf{Serie} & \textbf{ID} & \textbf{Weight} \\
\hline
Galactic Gecko Space Garage & AGkdTcHFasnsvH4ofrcK5FXasmvghnl1ZSBHcKGj5BrDDW & 0.094517 \\
Galactic Gecko Space Garage & 2XNg9ScN6vyMB7F5kdNZdwc6w7Fa3CD3Oeix5s5xnJbV4w & 0.150622 \\
Galactic Gecko Space Garage & 2K6yCDMWUNLMvRB55S1XaJymKUpKqzivQyerFVoeWP6 & 0.122612 \\
Galactic Gecko Space Garage & B7aemXNu4qA3dudis7H9am3KG96m6m3CqkSClcTpTF3 & 0.120258 \\
Galactic Gecko Space Garage & 2TXdfDLyrkCtm7Di1hmf6xbR9DHYjggpJSia8ojr9QYbL & 0.08115 \\
Galactic Gecko Space Garage & 39Be9GmugkMEwsAysdzmW3hmLVxnJ7BNHBnPfaBvk2jmv4 & 0.094866 \\
Galactic Gecko Space Garage & 5igAAxFmxwADkiMCzYD3yvPBxJYD5ELES1X3KBMMZsa4p & 0.123284 \\
Galactic Gecko Space Garage & Henzzr8C2kdkMetBct51wMJQ5Z9u1krRB2u4lcDUJEgti & 0.067053 \\
Galactic Gecko Space Garage & CZtembLM4J5jUWzYFpNWp2bHEqDml5Wzex2WwBleR & 0.102909 \\
Galactic Gecko Space Garage & 8TzDOJDecyaxkWAeedM9ZLX7RXiF4EeqasXAjQmq6EQ & 0.067017 \\
\hline
\end{tabular}
\end{table}
\begin{table}[H]
\centering
\caption{Portfolio for Degen Ape}
\label{tab:degen_ape_portfolio}
\begin{tabular}{|l|l|l|}
\hline
\textbf{Serie} & \textbf{ID} & \textbf{Weight} \\
\hline
Degen Ape & 99uc7fJh32vHwRwffi8Ncnrr3ByRZL3SQLvTfwyth73D & 0.180677 \\
Degen Ape & BfrSvmwhrJb1rA6uFHYRWCU4P6Ad8mHLEpC6G2YvN1JR3 & 0.048066 \\
Degen Ape & 4v7v7YTGD6JXSRubJggtN26zxAeK1bmMoakWCJjy2ef & 0.051087 \\
Degen Ape & PCjkYa9pH2HF26CJxgwc8EozadjuKoBgAv7r3bsrbfo & 0.132624 \\
Degen Ape & FN4JE2zpR4Y2kc88SVQXcm22FryTfynWiaPtg9BxHq7H & 0.108004 \\
Degen Ape & 8KYVXjxBV3fSnHg31xBYReu6ggz3Wzwr9jwUx4J2Abbn & 0.035049 \\
Degen Ape & 9nEqQri6E7Jf7z7d9Ag9AqGMfkcE4UFGkYYtcP25UKAM & 0.072588 \\
Degen Ape & 7FerQswjAYuhcadobW2jJ8RUWRb65cnWmm2X26w8hDi & 0.13676 \\
Degen Ape & 4EvdRYmBVWMaAGcu6T3Tioq6hmfrZh8E8bYbWxBfS31n & 0.107147 \\
Degen Ape & 62PNDR6e5RLk1RnZ7jHZSYW5vquBjuHlhqHgN85Gn3qR & 0.127996 \\
\hline
\end{tabular}
\end{table}
\begin{table}[H]
\centering
\caption{Portfolio for BLOOD. IO PASS}
\label{tab:blood_io_pass_portfolio}
\begin{tabular}{|l|l|l|}
\hline
\textbf{Serie} & \textbf{ID} & \textbf{Weight} \\
\hline
BLOOD.IO PASS & 549sCo6tUpJz8h84gAGN776nAPE5nvaRAna6ksfTMJ3x & 0.097675 \\
BLOOD.IO PASS & 7NFjK9c4XNknNxn7fwBVzLlkHCFsSTCz86mrjNeHRsQna & 0.05709 \\
BLOOD.IO PASS & xmozPJKJZXlhvug72hPjK2EqXiDD5hQGPYc26WgZjAB & 0.10367 \\
BLOOD.IO PASS & 9cHuTKYq1g1YigVDJQiauEzn8uPpGDSSgl1v9iAbiGz4Q & 0.075953 \\
BLOOD.IO PASS & xDmdEZQShcDmwN6mZG9A2wQ42DKVSKFzWrmARubryeVqK & 0.112238 \\
BLOOD.IO PASS & 7C7rV1ntUvTmhY8VvJzLj5KEsmsqcBFYkmE2O8v7Vmp1M & 0.088471 \\
BLOOD.IO PASS & HkbAm18SdfrtyCZ8nD9tQBNbid3zpDZSNckE7n4zDRY7S & 0.160822 \\
BLOOD.IO PASS & AQAMkzUu6GfKfKTEsLQoQWsJBFDeUE6ZGPhnNS12FRP8zc & 0.095984 \\
BLOOD.IO PASS & JAVVYg9zr44fLhUEe6tAhPQzyhJ9PwK2CXiDx4qhfmj & 0.097559 \\
BLOOD.IO PASS & 57GjCAtR5rJRspEQWdsT4cxJkUzSTLxpAPJqpfmpTwt & 0.114539 \\
\hline
\end{tabular}
\end{table}
\begin{table}[H]
\centering
\caption{Portfolio for Alpha Gardener}
\label{tab:alpha_gardener_portfolio}
\begin{tabular}{|l|l|l|}
\hline
\textbf{Serie} & \textbf{ID} & \textbf{Weight} \\
\hline
Alpha Gardener & 9NXbDTNwdK8WgAophPT92FnwkCQAueVBuYhpe4qcbZoWu & 0.065208 \\
Alpha Gardener & 3g8nJczivZbzB7jF4zuK9Lh2XKHvToS9RACBgN2jngTVZ & 0.073699 \\
Alpha Gardener & Bkb6tocSoC52UXeEkHcMVCkhj8eP9Ucgf111v4xzXGYy & 0.117214 \\
Alpha Gardener & 9xDV2wDFQThJrxvaQ1DtVcnooma3TueSxyC7rY3jJ7mNX & 0.12907 \\
Alpha Gardener & EvksKQboACmPkpPsVYjVTuCsV7BhtEBzBKDRarri8BRQ & 0.123471 \\
Alpha Gardener & 7WYvBbJuaNEMUeHGBzk37LSiyJB8X8zrosT5vOpcmx6wxd & 0.093729 \\
Alpha Gardener & GqVdwuzuYgFDebJ2TccV6AGsCSepPFShpVRquRxpPy8Mw7 & 0.097566 \\
Alpha Gardener & HJi5k2pd7uVS8gweAwuTmpthAbnzcCZpK6i d5BnKDQwK & 0.077987 \\
Alpha Gardener & 2V68ChrInBpU1 jvq jSqk iQJdmQc7C22o6ummmkPscBLcw & 0.064803 \\
Alpha Gardener & 6vhNbZecGUL6HXZ9habQFW4f2JBKTgpt7xt8Z6KQmJDZ & 0.157255 \\
\hline
\end{tabular}
\end{table}
\begin{table}[H]
\centering
\caption{Portfolio for Dappie Gang}
\label{tab:dappie_gang_portfolio}
\begin{tabular}{|l|l|l|}
\hline
\textbf{Serie} & \textbf{ID} & \textbf{Weight} \\
\hline
Dappie Gang & AGpTHKh7jQ88RieRru3CX6B3MAnLrWbaBXbiXLEWa & 0.033969 \\
Dappie Gang & Czboi98HQDqNEmgcuCZRRpg5iFVh4NeAXC4TLoiUo8sC & 0.150486 \\
Dappie Gang & 4NxRZFg5xqqr7aF25VEMtwogTNByGfJ7hwXwd3NFwXyv & 0.11618 \\
Dappie Gang & 3QQEmd3rDyHK4UDNwDmJhmMFfcvY4EcfBHAW4Bpxntml & 0.093857 \\
Dappie Gang & 9bQtR4E82DJW4v4JWkx2WSc4jCbdQ5y1P1Cdljfe64YJ & 0.099955 \\
Dappie Gang & zTFneSZ3DNAmnTRD3J43iV5iHq9Em7ky9R9Gbw2N51w & 0.06051 \\
Dappie Gang & 6MUX81btYzXdDZLeh7yW3bS1FG659rNYzyShzx6RrPFXZd & 0.103286 \\
Dappie Gang & GFXzggvG5ULusQaPZU n37dWAvpmLaJVPCSVQqt5KHCD6 & 0.130882 \\
Dappie Gang & E9E2zsLg3k1hWQGCfk88SBv7eAZrVzt2LGp9Uis2cAwgX & 0.053654 \\
Dappie Gang & 47kmKRsBUNuH5VFeRcmew8taGNNcmGAm6vGLin3WxUuXi & 0.12168 \\
\hline
\end{tabular}
\end{table}
\begin{table}[H]
\centering
\caption{Portfolio for Aurorian}
\label{tab:aurorian_portfolio}
\begin{tabular}{|l|l|l|}
\hline
\textbf{Serie} & \textbf{ID} & \textbf{Weight} \\
\hline
Aurorian & 4cfN0b1bvpTBJVj262xmaKoaN8ChkqetdzVGTzsNu6ex & 0.112044 \\
Aurorian & 2sEZextiyBMUNgXxsvedbztasgYmLRdd2jP72rWv4DoW & 0.109016 \\
Aurorian & UKs64RK3PpdLFgTP6moCiqu5azvbL2wkBSbPptzzrTz & 0.103812 \\
Aurorian & 2VABsSZzxPif4qjeSNNmovHnna33RWSNC19artKuS1CjJ & 0.122614 \\
Aurorian & CSuAGuyWVJjroJnxgaqGC7CukBU42BttmYzbYVUs7LAMj & 0.110863 \\
Aurorian & F1P6PjeA9cg5eiMAxEm8ECN43tDWkgGt4aHK2ATBZyBLF & 0.091849 \\
Aurorian & Hd2aJeeinf3PDLbxRPYr7Giz3FLZsAa2nXKWHDg1PN5Z & 0.080195 \\
Aurorian & 3LhvSNBziD24cWHDKgj nMJKvxOtxn2CrhuwZTYBosGMj & 0.120744 \\
Aurorian & 9tQ4odLB1LmLPwzzxw3UxRfNUFCmSG9SMLWDKdvcS27W5 & 0.082134 \\
Aurorian & Fxdi4VFSQ9QLrAhyZBKvrMgdAeKgTncKvawsVccnKeCs & 0.066729 \\
\hline
\end{tabular}
\end{table}
\begin{table}[H]
\centering
\caption{Portfolio for Ovo1}
\label{tab:ovo1_portfolio}
\begin{tabular}{|l|l|l|}
\hline
\textbf{Serie} & \textbf{ID} & \textbf{Weight} \\
\hline
Ovo1 & 8Z7rqyW1ysmMj6gRuFvJJC4eh4dtZegjaJYV18rfYfsC & 0.151064 \\
Ovo1 & CPDeCW579LqTWKacVJGUCGLxonJMyeXuJrFF6ZRefYtsx & 0.082613 \\
Ovo1 & 5uFF5DF9MecSyYDHQQQTVt3esX4BxDkY6NGXC35GyCFJ & 0.093908 \\
Ovo1 & 3NV6rgZSDmha5CukqW51bBo5JdHNH8SauLNkm2tFuJhpx & 0.071181 \\
Ovo1 & 8TGEfMB5EGUE6HV4tGhCJoZW8vJRzBLztf1eWUP4WhWM & 0.060748 \\
Ovo1 & A18kef6TT2XK8F89kSTMEhEnXY6SC9GxPswC6tyxxAq & 0.044942 \\
Ovo1 & 5mTFaoPBiSvCCNLACXJmDDpilrfnjbEZMT8Vv43mJ4LM & 0.098761 \\
Ovo1 & FDZfastuaqVeaBvFvrnDmAoZfeEekdcqz28seasGMZcpy & 0.154855 \\
Ovo1 & CkCYGWMqBTFvLD1WujrJV7kVNmNKvjd5UsV45JJnE & 0.106808 \\
Ovo1 & 6GntnwW54odvJN3HsaNalmLkn0XLhR7sq3RD6GGQNf3Gz & 0.13512 \\
\hline
\end{tabular}
\end{table}
\begin{table}[H]
\centering
\caption{Portfolio for Beast}
\label{tab:beast_portfolio}
\begin{tabular}{|l|l|l|}
\hline
\textbf{Serie} & \textbf{ID} & \textbf{Weight} \\
\hline
Beast & BNQ7gVRZuNZlJ97ekUrBGvpjvpnDwY1wLbCvhnfb9CJD & 0.110164 \\
Beast & B2uBRGr3E6Ngkw75mEbppiMUE3qzMcY81MQYFi1Qtt8T & 0.051987 \\
Beast & G3cDvmipcgLQWL8ichJWsdDDEgV2XpaGMz2sSK6Ump & 0.061447 \\
Beast & Fcq2a5PdoPCT7gQAba2YyGWgH8gv46NGZYEpJE5qJ1dj & 0.109936 \\
Beast & CdYWiLrqPoWMmrMBxETtruapE56SMVWNTJzq7AxnjM3G & 0.118709 \\
Beast & BhkXpwkrjgbC8RGlRRoxslrxutNTL2ehpmEBU5SYDx7G & 0.107245 \\
Beast & BUSUZ8cXBXPDUwNCzUSgDSgJFLHhW7RyWMJUuw55FHZax & 0.171164 \\
Beast & 913DduqicDK6kW5yJMLpFj1i7YzP8mjlhWWAGJP3LPFpq7 & 0.085823 \\
Beast & TyaCEYcfuCM4wcxyYTb66uJjN3j5f7iQpAibfyQqncsMA & 0.120421 \\
Beast & FyTSumanEb2nxvHeDnhCvOyJ3fnszhzGr8WWpR2wCg2i & 0.063102 \\
\hline
\end{tabular}
\end{table}
\begin{table}[H]
\centering
\caption{Portfolio for CHADS}
\label{tab:chads_portfolio}
\begin{tabular}{|l|l|l|}
\hline
\textbf{Serie} & \textbf{ID} & \textbf{Weight} \\
\hline
CHADS & ATgUWP5YZD6uW9XFqJTXsN2ZmNUrDzjpBT9XdaATxEoR & 0.1486 \\
CHADS & 56fWQA24932Y21jeFVGCpXUi1Qo5xxy9LCpwnEJ6Rvox & 0.10444 \\
CHADS & 4ggAkkp59je8tHkm9d4QxVh8fB6fwYCTHNNrDnn9uzdS9 & 0.117526 \\
CHADS & 2rdf7kkZK6ioyfwfyNc39agEnWrFHnXC5UfpcibG3rjLB & 0.094137 \\
CHADS & 756g7kXvb3rEto4hH3sXe1HZEdTVbJaqcP4CZY7ggBPY & 0.091999 \\
CHADS & E3Gq8ofgopobAnn8TqKo7bxnFB6shCTfEvSn69bTM2rP & 0.07625 \\
CHADS & 8D9wrbHfcybnStd498YCZ6QY7a5i6ez5icBUantZAkWP & 0.04402 \\
CHADS & DfbWUBJi91vuY23LwPERMJxrXSugt5e3DTAghmcnTjQN & 0.096069 \\
CHADS & DKAorJ52vyEtCLJUDa1AzjSmWaPc7zEiicHGW9RGdxjb & 0.146888 \\
CHADS & s16uuB6KAjHUoUX2rvuoYb8Y4urEstk7e9VDBTj6agQVT & 0.080072 \\
\hline
\end{tabular}
\end{table}
\begin{table}[H]
\centering
\caption{Portfolio for Asset Dasha}
\label{tab:asset_dasha_portfolio}
\begin{tabular}{|l|l|l|}
\hline
\textbf{Serie} & \textbf{ID} & \textbf{Weight} \\
\hline
Asset Dasha & FE68p8Ea8DxRwsiSe31atnG5KBqzxJERZsR247i4ACrm & 0.045206 \\
Asset Dasha & BTosgHu2faqFL2SbUim4e7dhdBmRlurdEFHs49V2ifLf y & 0.103462 \\
Asset Dasha & BdQMPmDz6DWmnQsyT6vJqeXY8KFkKH3wCVpeAuq4EVt & 0.0849 \\
Asset Dasha & ASHE1BLWaXz4amdDFw44cNLX1SYJbUbE8wYiSKeJP4b & 0.086867 \\
Asset Dasha & 6xjVoWkSFJ8YbnxBbCLNdJmAiR4poeQGumanJNPqxV5 & 0.11176 \\
Asset Dasha & BNttyKa5jiX49CjXNS5C8sPsaKU7dbfUKQKejV6GsQhQb & 0.100946 \\
Asset Dasha & 9kTemokN2dEazeVn4taAC2QzME1SF6jNhND4si6PGroB & 0.131874 \\
Asset Dasha & 9rhzuUZVpoKtJagFJvvBeL9KzgTu4xt7H75yLsk3v89k & 0.121194 \\
Asset Dasha & 2mWHtB8pCdvozigLpRrR7tvnMUDufQyWcPBstzcZodg & 0.101763 \\
Asset Dasha & 3rdlN6xtTfi7yqZFFPS8SmR2DyGCuWBQRKejM9TeBW & 0.10628 \\
\hline
\end{tabular}
\end{table}
\begin{table}[H]
\centering
\caption{Portfolio for SMB Gen3}
\label{tab:smb_gen3_portfolio}
\begin{tabular}{|l|l|l|}
\hline
\textbf{Serie} & \textbf{ID} & \textbf{Weight} \\
\hline
SMB Gen3 & 25PsVjDupteheargbY5DVFR8yiH1sCbMksk5vi vmJ9qu & 0.061138 \\
SMB Gen3 & G8G11AbgUWGN8kq3K3XtECpiBCnAugueruZHM6gfjVhe & 0.077803 \\
SMB Gen3 & Bpr9zdbzzV1cwgpxEJb7aBXJlWWwJkxSAWmPBG7QuqE2 & 0.082595 \\
SMB Gen3 & 9XMkaSXk1ebxs9ACp9u8WxkN7hhhozzyT4erCnvYBnqKH & 0.128079 \\
SMB Gen3 & 7HpQNNLsxNvWJc9RvaU29oY154nKEkvyv9EFrjFqN3yr & 0.123526 \\
SMB Gen3 & 4ckEyo8VstbMdqAmjorJaNCmxtpAEpAisEAnU9nE5yWi & 0.085796 \\
SMB Gen3 & 2PPicP4GKslY3C9Ryf4zA5DJnvco8XnCfePj5FscU6tT & 0.100306 \\
SMB Gen3 & 6q7YDMggh98Wdf4sMLPqChgCwBrH7uEvwPH1I bBR95UmP6 & 0.154229 \\
SMB Gen3 & 7WqMa5Zn47y5TX98YYWeto4nLaMhRFHLC2yU1ENFW4Tq & 0.07067 \\
SMB Gen3 & 6JvSgn8Whl1EymRNCAQdEEqFsjMgU69N8Qg9Ezzw6SJP aH & 0.109828 \\
\hline
\end{tabular}
\end{table}
\begin{table}[H]
\centering
\caption{Portfolio for Portals}
\label{tab:portals_portfolio}
\begin{tabular}{|l|l|l|}
\hline
\textbf{Serie} & \textbf{ID} & \textbf{Weight} \\
\hline
Portals & 6qVAxFDswMqi28ExKaPJAsAbPkyrgxFcwPR6gR9qtngGDF & 0.102339 \\
Portals & Evxm7bY5wgGh2vzaGgFuh8Attg2YlufnyecC5YNSzvnurUG & 0.08097 \\
Portals & 67dt7j43JoEhrTcwnYa5GnhvYfzn8jEkakffu8nusXfJv & 0.096435 \\
Portals & Fz7X12bewzbivXJbCZjz229QQH4Qdowt5jAMEyFZcBQm & 0.113009 \\
Portals & 24jc6YR7TY6chmHME3wFv7mrU4wS9ZaJk5d6tHyVYgUH & 0.105813 \\
Portals & Bods5gAga3BvmGVezzJjx6frbxDRzrnXvg55E9yxbSa2M & 0.109874 \\
Portals & 7dLmHgqtWz2XP3kQTXY2MaRKVEVYnJmgYozTT8SoqU7W & 0.15417 \\
Portals & Y4s9trWbCsQdWpmrzq5nYgm4mbC3yhecT869Z2Bfga & 0.106054 \\
Portals & 8iSMJCBbfiTpZeRuEFSngUcKztKvPMC9c8BmAsDra9Qrt & 0.053518 \\
Portals & 9h3GRBSfJo46CgPcMGAJBN4xToXB44B34JdI5n2cRzHNG & 0.078185 \\
\hline
\end{tabular}
\end{table}
\begin{table}[H]
\centering
\caption{Portfolio for Blocksmith Labs}
\label{tab:blocksmith_labs_portfolio}
\begin{tabular}{|l|l|l|}
\hline
\textbf{Serie} & \textbf{ID} & \textbf{Weight} \\
\hline
Blocksmith Labs & 8NFTVmahAHEmRdJBy1q9saGpJrAv2NQCPj9xktnWFiCj & 0.05109 \\
Blocksmith Labs & 4tndcDLKpFcKk1qBdDlqH8Koxrrsg3sxgT5s6BD4DYEt & 0.092024 \\
Blocksmith Labs & 9g4XUAQ9wBpEgJzyjwvRLihSbt1bF9dXinJScdCbE4s & 0.073231 \\
Blocksmith Labs & 6UjULCzMBUZf8KaVJH7vb59i jtceH9pdTdeZdwK5gZ1n & 0.122439 \\
Blocksmith Labs & FSYJzxFHddhxLvxYCAW8CkbUuBLx5PBFbAyi2VKLN55ye & 0.10808 \\
Blocksmith Labs & EvCjD129EhsZfGHzSK3kmD95gsP0PQeEW7ZkNizelfQ & 0.116293 \\
Blocksmith Labs & E8Eylscqoh6CNwRE9dfWbnHrQ1aU3p8buRfsP9pg6v4ae & 0.108617 \\
Blocksmith Labs & 7PF2csNs4KmLSta2FYgnKcEr6XMnoahS6x3Xu6aKgjf1 & 0.122095 \\
Blocksmith Labs & 13zNBFmveVRiocs4cXBrD2olWZ10CMfEusecS4ims9Z2d & 0.081388 \\
Blocksmith Labs & 5azJn2ECEiyRvGm1Ks3vc4zWDdfJaJKMaNR8kzNgUYv & 0.124742 \\
\hline
\end{tabular}
\end{table}
\begin{table}[H]
\centering
\caption{Portfolio for Ticket}
\label{tab:ticket_portfolio}
\begin{tabular}{|l|l|l|}
\hline
\textbf{Serie} & \textbf{ID} & \textbf{Weight} \\
\hline
Ticket & BDj45c9yAFAsZQPF7VwJ5GNmCU5wEBkgHHAky4NP9Bw & 0.111737 \\
Ticket & 3yuyuhdo8J9YhMnztRW5a jgtZVmroM7VkvPsyNLAvm5B3L & 0.08276 \\
Ticket & 2XRsLKUfaPETuBJeSo5n8MUmLmHPzBYYwY6cDtbmDGFU3 & 0.120411 \\
Ticket & 5kDJ8qjpQwNaz4eMPwMXGHP182dcdBZvQfByQArFM1u9 & 0.114333 \\
Ticket & 7KVznzkoGjfnAnbgUXU87gEmr51SNwcYptkejb2cRzSx & 0.090501 \\
Ticket & 4WS22coX3UNG LrNEcStokgxJxUdWefmJu92SU RuTaXb9 & 0.114754 \\
Ticket & 8mrFUDSSFSVRaeR4c9FTYkmAqgHiVBSU124mrUJ2ok54Rm & 0.125037 \\
Ticket & 6SpZp3YsGp83YhPL8F jmm9KBuCbZinNinXwhnbgoNVR9 & 0.090018 \\
Ticket & Gbh6hVRunS6R1s8uYwjnKglBmnZLK7vzDxdvaJymZNj & 0.078877 \\
Ticket & E8VEVaGHytUh6o2Cxo3Lnx1UcbBAX1pkdXUnMtJbshm & 0.071573 \\
\hline
\end{tabular}
\end{table}
\begin{table}[H]
\centering
\caption{Portfolio for Clanynosaurz: Call of Saga}
\label{tab:clanynosaurz_call_of_saga_portfolio}
\begin{tabular}{|l|l|l|}
\hline
\textbf{Serie} & \textbf{ID} & \textbf{Weight} \\
\hline
Clanynosaurz: Call of Saga & G6kaAepM6bocXC5aYzFBCLqwinBQaRg2V64CcmXYvqgCj & 0.158797 \\
Clanynosaurz: Call of Saga & 2d3xnYXiqx9cdrYbZ5qYlDbko6NjBgePashnzEWykMCZMD & 0.094572 \\
Clanynosaurz: Call of Saga & 40KB7Hxa4HF7xbpbrHxnPLMlRDLTmqMFuYDKgeKZLde & 0.082894 \\
Clanynosaurz: Call of Saga & U1BfrR5tBKFPKsXnFtg3YJDJjEnvEsouIn2CM5BVEOde & 0.120413 \\
Clanynosaurz: Call of Saga & 9QTmsTVE8s3gUjdYNNMwZ2sUjfJe5pMV357ozK5AnthKK & 0.081728 \\
Clanynosaurz: Call of Saga & 8aCM986Lw5gbpxbToKxrAX7fFAv9avxFKQYtsBuCoCpJ & 0.088636 \\
Clanynosaurz: Call of Saga & 7k3bcBtmL6clgxZTfidbBWYTVEnnxNa6fX12eCxCvAYk & 0.122418 \\
Clanynosaurz: Call of Saga & 95amPSZKg4teTYaA2bMw43Pwpm5S5EwpdLF65jqp2C6P & 0.113161 \\
Clanynosaurz: Call of Saga & 80qnHAXPpyzWpAWnQ6dViJAcLBhQTnrHn3aw6CbkXtJia & 0.074642 \\
Clanynosaurz: Call of Saga & HZfTnm2YUdNRiTGrnWlVQWhQRqHVspYSkT87f2wFsR & 0.08074 \\
\hline
\end{tabular}
\end{table}
\begin{table}[H]
\centering
\caption{Portfolio for Bozo Collective}
\label{tab:bozo_collective_portfolio}
\begin{tabular}{|l|l|l|}
\hline
\textbf{Serie} & \textbf{ID} & \textbf{Weight} \\
\hline
Bozo Collective & FBrcQHrSBjJ1F3gYmqkczjx2bJc1fjaKMtsTc4HMVsz & 0.072152 \\
Bozo Collective & 6LoHYLUA5mcBAsdoeqiNwWsVB2EBESq4JQkzLFNV9vCe & 0.12444 \\
Bozo Collective & GS34iNEgBm5AX77NzUg7vQrpqYmBgawhAMUjQk3iJKkx & 0.146667 \\
Bozo Collective & 2aRigguvINVM7k972F36WxSxxiQ2uib1vFWY6zpK14bo & 0.115924 \\
Bozo Collective & C26tzgg8RCoBiVj7ZbfZj6wJ5JEfS7hZLzoFh6TwCJrG7A & 0.141768 \\
Bozo Collective & Dkg4eW2xiZB668c0FaWSbXR4jS3dfdnUcaCS6GQkmYrc4 & 0.085664 \\
Bozo Collective & 3mYkfTglkjCsN6tnWzSbPbQQ9UrTZFMPA48rE8TXPrssr & 0.048039 \\
Bozo Collective & AMkaZvQF6A4r6Ss2ynr1132Sg3GXcEkL5WSJ5hoFHf8g & 0.130551 \\
Bozo Collective & 4BbCxZwwpU3HomUxJnfNDLkXrodnIAawJYiunDlEyFZ & 0.060608 \\
Bozo Collective & 5jyucVyZn23rcP92exedbCD8EledmqG334ah9iYrQfw & 0.074188 \\
\hline
\end{tabular}
\end{table}
\begin{table}[H]
\centering
\caption{Portfolio for Meekolony Pass}
\label{tab:meekolony_pass_portfolio}
\begin{tabular}{|l|l|l|}
\hline
\textbf{Serie} & \textbf{ID} & \textbf{Weight} \\
\hline
Meekolony Pass & HzKbheXESqMu94MPSVrktFQ6V7spSpiIIUQsVojLW5a3x & 0.079024 \\
Meekolony Pass & 4M6hvDq5PpFLRiuLnT4EQ1SVuyuYX3YsWtW47aoGnGq4 & 0.086602 \\
Meekolony Pass & CFzqYkhX8Q4PgE3sp3VWLaFdSpaVscTuiA4KaZQJWAL & 0.077212 \\
Meekolony Pass & 3Xzw2jUWS944JHGjexcCyZNoweaCDPZtuP j7ynYB i8psd & 0.110042 \\
Meekolony Pass & ahRS629gWTrZ2DhqWexJDecxeJKVMoQYoLasS6tguGhZ & 0.125243 \\
Meekolony Pass & 3Yh1T1hLUzPKleH9TLHMdeJ4XjMp7fSmnD9rBAi4Mzxj & 0.08987 \\
Meekolony Pass & 2XjNARRg4AFFf8xSVgRwqR2uKuahmn9nhnaKvz2KUiBo & 0.162573 \\
Meekolony Pass & CgVJR4n6vzmGvN97xbuaGcyS5ChErzYcjnracX3Sqksx & 0.085186 \\
Meekolony Pass & Caw5MSnybESpLyMXpPDcx3nW4oyMNwiLysbNWvtJzLF9 & 0.079899 \\
Meekolony Pass & 97mgBBTLWjJffkmLAGbZRRqaxptpmlMntcrC97aUgvF2ux & 0.104349 \\
\hline
\end{tabular}
\end{table}
\begin{table}[H]
\centering
\caption{Portfolio for Fred NFT}
\label{tab:fred_nft_portfolio}
\begin{tabular}{|l|l|l|}
\hline
\textbf{Serie} & \textbf{ID} & \textbf{Weight} \\
\hline
Fred NFT & 8zHy9RSVaJGZYJcHm6BoJDgFta6bvv9NdnwZY7EieGvg & 0.106062 \\
Fred NFT & GhybD6Rq42s4aLPd4s4xpV9aXxfJoJvtV4AS2ZqvyRzSoYP & 0.067892 \\
Fred NFT & 5N5PjssakmtREtbW1fnyoGF4YbRApDSdy4uNwcD8ZYLm & 0.137533 \\
Fred NFT & 63uJnFi5j7nomYeh6ZChmzyhHLyBHOMnFCooMX9zRR4Y3 & 0.13922 \\
Fred NFT & EsvMarcoW4SSBk4jJXt7YhYd9R6MFdfdXYbhLpVYvve3HkgF & 0.100033 \\
Fred NFT & H4QhssSPEMCh5r7o38Xmc9TUmgGL5tKqavKfyXxv4LDer & 0.004196 \\
Fred NFT & 7rYXG8HRkc jd48mmr2pXmgdZXxoFeHikm7sJfbQJDBVe & 0.131646 \\
Fred NFT & 36yuUZjfmv18Lq2qmnQnSgtE9CAoKhrTATRRV56rom71W & 0.091758 \\
Fred NFT & 2g7W9m217j3Lat91GCJGRZvDqaw8gYUCUWuusuMeG32Pa & 0.097866 \\
Fred NFT & 6RSKhd341aDPsh9phhfNu43T3kSRY8ujPiLQsSz6FkD & 0.126775 \\
\hline
\end{tabular}
\end{table}
\begin{table}[H]
\centering
\caption{Portfolio for The Heist}
\label{tab:the_heist_portfolio}
\begin{tabular}{|l|l|l|}
\hline
\textbf{Serie} & \textbf{ID} & \textbf{Weight} \\
\hline
The Heist & 48WT7ea5YdspkY9NJGgWUsdRHRGloRT8PNdrbfZ1YarjP & 0.083681 \\
The Heist & Cp9H7ohMFHSuktoErY46sVqkQQ7NzSCaeADZYhHJ2KY4t5 & 0.106894 \\
The Heist & B9Me9BVie4NUDZFJGhmMn7qaGQAPkQzrSQXUWPlJcbG9c6 & 0.158856 \\
The Heist & 5TuNJr5JaEalSgluYdhbWz48hky8wpxrBFuJr3Qs1R4h & 0.108851 \\
The Heist & 5f8npl4FaikHTYo26d8Dq1cwtfH2WTJz36DCmdYDCdcg3 & 0.0682 \\
The Heist & 58eYsF3xWneka6n8eusDcmyce8u5geyxQde3t3PPfk5A & 0.075521 \\
The Heist & 6bFrEQSP4QrxSLuHwahdwc j75hzwWYZkmybAsmBvYxNJ & 0.133831 \\
The Heist & CDWwM9pcwk76eTm5FzAvZyNhTfSahTwcWLJpgpssaamxC & 0.078515 \\
The Heist & 85p3b17gCvJPVS4jCJq4ukNbxsaCdMARSvh7MTosSHXk & 0.09983 \\
The Heist & EWkDrxSFO2KqQXPsaACynBQpAeVjKegtWCbwBgKC7mYhl iG & 0.085821 \\
\hline
\end{tabular}
\end{table}
\begin{table}[H]
\centering
\caption{Portfolio for Degen Fat Cat}
\label{tab:degen_fat_cat_portfolio}
\begin{tabular}{|l|l|l|}
\hline
\textbf{Serie} & \textbf{ID} & \textbf{Weight} \\
\hline
Degen Fat Cat & GMkadUhh8Jrb4Jnzvonr7SH53ecJ84ZtgoK8E2EkHdJJ & 0.069619 \\
Degen Fat Cat & 796ANQaMGutbLUFmZjsMb6AH29PWS H4UMT27WM7dEs8r & 0.090749 \\
Degen Fat Cat & AeDeqXUW8o j92bP2mmBpPqAzU5wwwivyoUwzgMFFEhDR & 0.101542 \\
Degen Fat Cat & EVAyhbjge5kwDMDLdgPSSchDgwvSbrerFAvqBqLZEtJ2 & 0.153879 \\
Degen Fat Cat & DuzbXWRtacZB7jwKgtBKo3HREMjSdQpYuBgxuEDX5rJV & 0.057552 \\
Degen Fat Cat & Dbxp8eL2bVU54zwtRSwK9gscZ4UDwMCmn76ix6BAys & 0.105039 \\
Degen Fat Cat & B4Mf1EbqXiA8gYQMg2wo76Cqc8tbfELNp7vhbMKbzXm & 0.108898 \\
Degen Fat Cat & 2pqVYgn64uYQT3YrVfgCiPHvTbEqWiluHzkoXKg4sxQDB & 0.104243 \\
Degen Fat Cat & 5Px4XzMohmdxBn8xiFY3JZZLfA9K1RwJS3CPM9bzJbY & 0.092671 \\
Degen Fat Cat & 9Eug6CdjpBVJXGjbRKNLLZJiqk4hlnCGynYjrXfqKazP & 0.115807 \\
\hline
\end{tabular}
\end{table}
\begin{table}[H]
\centering
\caption{Portfolio for Honeyland}
\label{tab:honeyland_portfolio}
\begin{tabular}{|l|l|l|}
\hline
\textbf{Serie} & \textbf{ID} & \textbf{Weight} \\
\hline
Honeyland & GEDFPuf71HPVExRkT1FNVyLN2BNc8wiXmkpNqKMSAMRY & 0.105055 \\
Honeyland & 7gVD4CRGjzz9JvemGNrAYf7ZzuKM2eVxpPZJCpKWYSjqQ & 0.085069 \\
Honeyland & x7DuSH6ELCNHvW8AZF2hJgUJCcwfyAA6t6SukyiJbcK & 0.121546 \\
Honeyland & 3kERR3nBi8GScUa7jU5qF6ECBAKAYf5cXVekFuJyFdhe & 0.132242 \\
Honeyland & Bu9jRDxrZp8kUSGq3N9dQPdUDnVN1JeoDWjqfh1BN3aX & 0.102705 \\
Honeyland & 9dDaShykmiiRrnmppmws6E9FhmG6G4xdT3rvB2WJCfprV & 0.068198 \\
Honeyland & hk63aPXJHsmF4YY8xHUMF8PWMWrPFN6wHdEjpuiD9nF & 0.073581 \\
Honeyland & 5gSY5uXKK1szRaa5uGu8GuEgJEcPS47mibGvwjCPWrRUg & 0.084763 \\
Honeyland & 9PZrn9BuFS8P54Pq52oaw744GuRa5gv8FaEDWGYERjbAk & 0.092272 \\
Honeyland & 3lhtMnuPPsKAo2n9NWkas2EvpSjpfvF4NaYE9CFCKeGHb8CX6 & 0.13457 \\
\hline
\end{tabular}
\end{table}
\begin{table}[H]
\centering
\caption{Portfolio for Frog}
\label{tab:frog_portfolio}
\begin{tabular}{|l|l|l|}
\hline
\textbf{Serie} & \textbf{ID} & \textbf{Weight} \\
\hline
Frog & FhbREJcnLNzzTvnMjCKtsqmpgtgdEaCP2HuWlexfuhzQ & 0.127664 \\
Frog & 9A8ssJcthuBE3k8Ltq415oPUQC9b96SFAR3fcVVPkPYlW & 0.020927 \\
Frog & 93TAMnXfV53Lj6dLBuioPoPPDBruHbxQQuDD9VCVZzc3R & 0.098313 \\
Frog & DXXawpwGCNVVuPBy7eRef3zPK5gDTONRrnkGvECWDjtv & 0.162892 \\
Frog & DppnVJaE1DLoeblc36BJNw5CFG2qwdjAGYJFCIAiFwz3w & 0.15539 \\
Frog & 7Cv26pcCRFNHBw146mSCYhjbmqP7GsuKpg5NriL4WJXy & 0.125189 \\
Frog & 699yuLYDqnPggf4x49Nvagy2TtVNwx6bJF jBjN6k5EMMw & 0.043736 \\
Frog & 6H9hrunVJN9DWApUzfNx9WxBteTR73T5kufqQYe81AZv & 0.012197 \\
Frog & 4bumKdc8vch9CMZ3f6J69TtQUQCFi1Nnw1IKX8NeSTmR & 0.131742 \\
Frog & 91sBrPS8HgLrzm5HEp4EzmaUUmUpkZTh8pk5HiHrXb & 0.121949 \\
\hline
\end{tabular}
\end{table}
\begin{table}[H]
\centering
\caption{Portfolio for STEPN}
\label{tab:stepn_portfolio}
\begin{tabular}{|l|l|l|}
\hline
\textbf{Serie} & \textbf{ID} & \textbf{Weight} \\
\hline
STEPN & 86ByhDLxghHrzV9rDLB5hKoc5jvC8n4fDb jo77HYTSEDC & 0.091218 \\
STEPN & GKN1z5KvdNhNh492sJgNDA2er6o1CDptmpRrWCT7nR2gf & 0.12483 \\
STEPN & HmvnTJ8q7sQ2CJy2nTQzqigfvSCexf6Bzrpqi8DCgsAe & 0.144232 \\
STEPN & FgKEmaqrRGUCTA9z97JJF69bK4AkmC6481QZJ8biptCC5 & 0.085015 \\
STEPN & AoC7StryrEFaeomiK4AZ2ASZqFQ1ExuP JrrZuPU6p2Nz & 0.091028 \\
STEPN & 4PEtLXZztTAZrzac3rdLAu8rJpDwHS9LYBcTFSq6DUDX & 0.090991 \\
STEPN & 2b84fXd7aJp68RccgAKjaeo14JognSCVTpvXh7KPme5 & 0.060749 \\
STEPN & EuuvBVQT4MoSByDGcSt9HCM1smdny99Hx3c68g9YdGFz5 & 0.081377 \\
STEPN & 6c9GryitdcJeJEe9uv5NnZ11FyNFZA3NpQg61A75F4XA & 0.110797 \\
STEPN & 3nWMMNLoa jphaUiEkbJ16LuFXqnsX4SWdWiFzbf54IHgdm & 0.119762 \\
\hline
\end{tabular}
\end{table}
\begin{table}[H]
\centering
\caption{Portfolio for Cega Super Sonics}
\label{tab:cega_super_sonics_portfolio}
\begin{tabular}{|l|l|l|}
\hline
\textbf{Serie} & \textbf{ID} & \textbf{Weight} \\
\hline
Cega Super Sonics & AM5CSavePtu393 ji8wQFxXYBWUt6icndwmxnKn8FsgTSD & 0.079342 \\
Cega Super Sonics & 4DqbcVB8BzMrv4v4ag8eXrRx945iyMGezMtTjwksskoOn & 0.128018 \\
Cega Super Sonics & 6eWca89SEi791mLRiHEYfc18o4KW4WKZZqg9JntEuWiP & 0.092108 \\
Cega Super Sonics & ABL8CRZMP2mk185moZvkoV1kHgriPvooyKL3y9RSEXLX & 0.172269 \\
Cega Super Sonics & B9X4DhCvH1hmWYFzFy2FJhtbjmvtzzrsJmr29tSUwBCMK & 0.088943 \\
Cega Super Sonics & GPMka2V8PWQwZcgaPQKiz72TR4jXCjfJsHyonJgQYBZf2 & 0.088631 \\
Cega Super Sonics & Gaa47qyFkhD5LM8xRSgvzrxrjl8mk2KBq6RBYcn9Jt5HU & 0.088943 \\
Cega Super Sonics & 6wiY5pEyiUsUrsrrh2DRYtn3yScKfehcAPRCTLnHmfNF2 & 0.092342 \\
Cega Super Sonics & 2nkibSROtMGdsTBn3aXHhrFGaYAYASBgqAYesGWbLnF & 0.05188 \\
Cega Super Sonics & bbxEtSMeG26MGagh5Sh5caS9V7sVJ8vrN4V5gzesYeqS4ag & 0.117037 \\
\hline
\end{tabular}
\end{table}
\begin{table}[H]
\centering
\caption{Portfolio for TYR}
\label{tab:tyr_portfolio}
\begin{tabular}{|l|l|l|}
\hline
\textbf{Serie} & \textbf{ID} & \textbf{Weight} \\
\hline
TYR & Bai8t8SPwExtJwndid35LbW3YNS9WJgk9KkTCAAhBotrU & 0.093321 \\
TYR & 2TdjXZYzogtVHfFxYxucjNJ25xXGohfawtRaBaQFXcQF & 0.076138 \\
TYR & 7BLNnALEoezUzs9c28Lqni1kU8T4RnH471wX81Wm6vJp & 0.078318 \\
TYR & BDEulscFTuWQQ9z8VjmAKjFr2eyuvuvAvBXefsjtduToVYG & 0.138707 \\
TYR & 6rnERRj3iviTpoaqFfjkbKDigxnSh7gr5zVTjuvwHR5mZ & 0.115078 \\
TYR & Gf7EkNDXibWmqiJazKA5ZDGozmRPLGZEhoVao tuoEZ7i & 0.088054 \\
TYR & pCLhwZnbSwrcCSXtDubCa87RbkrvXhW83yoQ1s9qCqQ & 0.063497 \\
TYR & ExgG93CX2LgMSKnpBhoSboQm3yCjFj1wLVPLMWwzC2eL7JW & 0.125263 \\
TYR & 51WMe9JVxhifa9oLZcS1M89aqhs12TaZikNxHk2KcoZ9 & 0.071445 \\
TYR & 7BhUj41aoyrqaQy36Gc2vEDFzqel9QemgG2d4kjgNFd6cb39 & 0.15018 \\
\hline
\end{tabular}
\end{table}
\end{document}